\documentclass[11pt,a4paper]{article}
 \pdfoutput=1
\usepackage{jcappub}

\usepackage{amsmath,amssymb}
\usepackage{graphicx}
\usepackage{dcolumn}
\usepackage{bm,color}
\usepackage{hyperref}
\hypersetup{
    colorlinks = true,
    allcolors = blue
}
\usepackage{amssymb,float}
\usepackage{amsmath}
\DeclareMathOperator{\sech}{sech}

\newcommand{\udt}[3]{#1^{#2}_{\phantom{#2}#3}}
\newcommand{\udut}[4]{#1^{#2\phantom{#3}#4}_{\phantom{#2}#3\phantom{#4}}}

\newcommand{\dut}[3]{#1_{#2}^{\phantom{#2}#3}}
\newcommand{\dudt}[4]{#1_{#2\phantom{#3}#4}^{\phantom{#2}#3}}

\title{\boldmath Testing the violation of the equivalence  principle in the 
electromagnetic sector and its consequences in \texorpdfstring{$f(T)$}{ft} 
gravity}

\author[a,b]{Jackson Levi Said,}
\author[a,c]{Jurgen Mifsud,}
\author[c]{David Parkinson,}
\author[d,e,f]{Emmanuel N. Saridakis,}
\author[g]{Joseph Sultana,}
\author[a,b,h]{and Kristian Zarb Adami}

\affiliation[a]{Institute of Space Sciences and Astronomy, University of Malta, 
Msida, MSD 2080, Malta}
\affiliation[b]{Department of Physics, University of Malta, Msida, MSD 2080, 
Malta}
\affiliation[c]{Korea Astronomy and Space Science Institute, 776 
Daedeokdae--ro, Yuseong--gu, Daejeon 34055, Republic of Korea}
\affiliation[d]{Department of Physics, National Technical University of Athens,  
Zografou Campus GR 157 73, Athens, Greece}
\affiliation[e]{National Observatory of Athens, Lofos Nymfon, 11852 Athens, 
Greece}
\affiliation[f]{Department of Astronomy, School of Physical Sciences, 
University of Science and Technology of China, Hefei 230026, P.R. China}
\affiliation[g]{Department of Mathematics, University of Malta, Msida, MSD 2080, Malta}
\affiliation[h]{Department of Physics, University of Oxford, Denys Wilkinson 
Building, Keble Road, Oxford OX1 3RH, UK}

\emailAdd{jackson.said@um.edu.mt}
\emailAdd{jurgen.mifsud@um.edu.mt}
\emailAdd{jurgenmifsud@kasi.re.kr}
\emailAdd{davidparkinson@kasi.re.kr}
\emailAdd{msaridak@phys.uoa.gr}
\emailAdd{joseph.sultana@um.edu.mt}
\emailAdd{kristian.zarb-adami@um.edu.mt}

\abstract{A violation of the distance--duality relation is directly linked with 
a temporal variation of the electromagnetic fine--structure constant. We consider a number of well--studied $f(T)$ gravity models and 
we revise the theoretical prediction of
their corresponding induced violation 
of the distance--duality relationship. We further  
extract constraints on
the involved model parameters 
through fine--structure constant variation data, alongside with supernovae data, and Hubble parameter measurements. Moreover, we constrain the evolution of the effective $f(T)$ gravitational constant. Finally, we compare with revised constraints on the phenomenological parametrisations of the violation of the 
equivalence principle in the electromagnetic sector.}

\begin{document}

\maketitle
\flushbottom

\section{\label{sec:intro}Introduction}

It is well-known that the $\Lambda$CDM cosmological model is evidenced by 
overwhelming successes in describing the Universe at all scales where 
observations can be made \cite{misner1973gravitation,Clifton:2011jh}. This is 
achieved by considering an extra  cold dark matter sector which can produce 
stable galaxies and clusters thereof \cite{Baudis:2016qwx,Bertone:2004pz}, while 
the late--time cosmic acceleration can be described through the cosmological 
constant. However, despite extraordinary efforts, the model still retains 
numerous questions that still appear insurmountable in this regard 
\cite{RevModPhys.61.1}.

On the other hand, the efficiency of the $\Lambda$CDM paradigm in explaining 
precision cosmology observations has been called into question in recent years. 
In this respect, the core criticism of the $\Lambda$CDM model appears primarily 
through the so-called $H_0$ tension problem which quantifies the inconsistency 
between the $\Lambda$CDM predicted value of $H_0$ from the measurements of the 
anisotropies of the cosmic microwave background (CMB) and its reported value 
from local observations. The problem first appeared through measurements by the 
\textit{Planck} Collaboration \cite{Aghanim:2018eyx,Ade:2015xua}, but has since 
been confirmed to a greater degree by the strong lensing measurements from the 
H0LiCOW collaboration \cite{Wong:2019kwg}. In the interim period, measurements 
on the tip of the red giant branch (TRGB, Carnegie-Chicago Hubble Program) have 
yielded a lower $H_0$ tension \cite{Freedman:2019jwv}. The problem may be 
further illuminated by future observations from gravitational wave astronomy 
such as through the LISA mission \cite{Baker:2019nia,2017arXiv170200786A} 
similar to the work carried out in Refs. \cite{Graef:2018fzu,Abbott:2017xzu}.

There have been a plethora of theories in recent years that attempt 
to describe the disparate phenomena that make up observational cosmology 
\cite{Clifton:2011jh,Capozziello:2011et}. Collectively, these theories are 
primarily extensions of General Relativity (GR) in that they consider gravity 
through the prism of curvature by means of the Levi-Civita connection. The most 
popular of these theories is standard $f(\mathring{R})$ gravity 
\cite{DeFelice:2010aj,Nojiri:2010wj,Capozziello:2011et} (over-circles are used 
throughout to refer to quantities that are calculated using the Levi-Civita 
connection) which is a fourth-order theory that directly generalises the 
Einstein-Hilbert action.

In this work, we consider the possibility of gravity being a manifestation of 
torsion through the Weitzenb\"{o}ck connection \cite{Weitzenbock1923}. 
Teleparallel gravity (TG) is the body of theories that are based on the 
Weitzenb\"{o}ck connection \cite{Cai:2015emx,Krssak:2018ywd}. While the 
Levi-Civita connection is torsion-less, the Weitzenb\"{o}ck connection is 
curvature-less, and both satisfy the metricity condition. The Einstein-Hilbert 
action relies on a Lagrangian that is simply constructed by the standard Ricci 
scalar, $\mathring{R}$, while TG can produce identical dynamical equations for a 
Lagrangian that consists only of the torsion scalar, $T$. This is the so-called 
\textit{Teleparallel equivalent of General Relativity} (TEGR), and differs from 
GR only at the level of Lagrangian by a total divergence quantity, $B$ (boundary 
term). The boundary term embodies the fourth-order corrections which arise to 
have a covariant theory (due to the second-order derivatives in the 
Einstein-Hilbert action). The importance of this property is that extensions to 
TEGR will differ from their Levi-Civita connection counterparts. 
Moreover, TG has a number of interesting properties such as its similarity to 
Yang--Mills theory \cite{Aldrovandi:2013wha} giving it an added particle physics 
dimension, its potential to define a gravitational energy-momentum tensor 
\cite{Blixt:2018znp,Blixt:2019mkt}, and that it is more regular than GR in that 
it does not require the introduction of a Gibbons--Hawking--York boundary term 
in order to produce a well-defined Hamiltonian formulation \cite{Cai:2015emx}. 
As an aside, the theory can be constructed even without the weak equivalence 
principle (but definitely it can satisfy it if needed) unlike GR 
\cite{Aldrovandi:2004fy}.

Following the same reasoning as  $f(\mathring{R})$ gravity 
\cite{DeFelice:2010aj,Nojiri:2010wj,Capozziello:2011et}, the TEGR Lagrangian 
can be arbitrarily generalised to produce $f(T)$ theory 
\cite{Ferraro:2006jd,Ferraro:2008ey,Bengochea:2008gz,Linder:2010py,Chen:2010va} 
which is a generally second-order theory of gravity. This last point is a result 
of a weaken Lovelock theorem in TG 
\cite{Lovelock:1971yv,Gonzalez:2015sha,Bahamonde:2019shr} which emerges due to 
the absence of the boundary term. A number of $f(T)$ gravity models have shown 
promising results in the cosmological regime 
\cite{Cai:2015emx,Nesseris:2013jea,Farrugia:2016qqe}, as well as in galactic 
\cite{Finch:2018gkh} and solar system 
\cite{Farrugia:2016xcw,Iorio:2012cm,Ruggiero:2015oka,Iorio:2016sqy,Deng:2018ncg}
 scale physics. The boundary term can also be included in this generalisation 
to produce $f(T,B)$ gravity 
\cite{Bahamonde:2015zma,Capozziello:2018qcp,Bahamonde:2016grb,
Paliathanasis:2017flf,Farrugia:2018gyz,Bahamonde:2016cul,Bahamonde:2016cul,
Wright:2016ayu}. In this latter case, the model produces a general framework in 
which $f(\mathring{R})$ gravity forms one subclass of possible models.

Another important aspect of any potential proposal  for a modified theory of 
gravity is its Einstein frame features. In many cases, the Einstein frame is 
obtained through a conformal transformation which leaves the electromagnetic 
action invariant (due to the conformal invariance of that action) 
\cite{NOJIRI_2007}. However, $f(T)$ gravity cannot be written in the Einstein 
frame by taking conformal transformations. In fact, conformal transformations 
produce an extra term in which the conformal scalar field and the torsional 
contribution are coupled \cite{Wright:2016ayu}. Conformal transformations lead 
to the Einstein frame only in $f(T,B)$ gravity in the limit in which 
$f(\mathring{R})$ gravity is reproduced. It is also the case that disformal 
transformations cannot either lead to the Einstein frame in $f(T)$ gravity 
\cite{Hohmann:2019gmt,Golovnev:2019kcf}, which implies that if an Einstein 
frame exists then it may produce a non-vanishing coupling to the 
electromagnetic sector.

The series of  works in Refs. 
\cite{Minazzoli:2013bva,Hees:2014lfa,Minazzoli:2014xua} consider the possibility 
of a violation of the Einstein equivalence principle (EEP), a cornerstone of GR, 
through the appearance of a coupling parameter between the scalar field that 
transforms the gravitational action to its Einstein frame, and the matter 
fields. On the other hand, in Refs. \cite{Brax:2012gr,Minazzoli:2015fwm} the 
violation of the EEP is considered through the presence of quantum effects such 
as the coupling of heavy fermions to photons. This may be the source of the 
potential violation of the EEP in TG.

It is well--known that a nonminimal multiplicative  coupling between a scalar 
field and matter fields would break the EEP, and would further lead to the 
variation of fundamental constants of Nature \cite{Hees:2014lfa}. For instance, 
a scalar field coupling with the electromagnetic Lagrangian would lead to a 
variation of the fine--structure constant, or Sommerfeld's constant, which 
characterises the strength of the electromagnetic field and appears as a 
coupling constant in the electromagnetic action. A variation in the fundamental 
constants of Nature \cite{2001RPPh...64.1191F}, which could be conservatively 
defined as those theoretical free parameters that could not be calculated with 
our present knowledge of physics, has been a long--established intriguing 
question \cite{Dirac:1937ti,Dirac:1938mt} with pertinent consequences for 
fundamental physics and cosmology (see, for instance, Refs. 
\cite{Uzan:2002vq,Uzan:2010pm,Martins:2017yxk}). Interestingly, when Dirac's 
numerological principle \cite{Dirac:1937ti,Dirac:1938mt} was encapsulated in a 
field--theoretical framework, this led to the birth of the 
Jordan--Fierz--Brans--Dicke scalar--tensor theory of gravitation 
\cite{jordan1937physikalischen,Fierz:1956zz,Brans:1961sx}. Moreover, in theories 
with additional space--time dimensions, fundamental constants of Nature are only 
effective quantities and are related to the true constants via characteristic 
sizes of extra dimensions \cite{Barrow:1987sr}. Such paradigms include, for 
instance, Kaluza--Klein models \cite{Chodos:1979vk,Barrow:1987sr}, superstring 
theories \cite{Wu:1986ac,Barrow:1987sr} and brane world models 
\cite{Kiritsis:1999tx}. 

A number  of theoretical models have been proposed in order to explore the 
possibility of a dynamical fine--structure constant $\alpha\equiv e^2/\hbar c$. 
These models have been primarily formulated as Lagrangian theories with explicit 
variation of the velocity of light $c$ \cite{Barrow:1999is,Moffat:2001sf}, or of 
the charge on the electron $e$ \cite{Bekenstein:1982eu,Livio:1998pp}. 
The former class of models are also known as varying speed of light theories 
\cite{Moffat:1992ud,Barrow:1999is,Magueijo:2003gj}, and have also been studied 
in the context of inflationary cosmology 
\cite{Albrecht:1998ir,Clayton:1998hv,Barrow:1999is}. The latter models are 
commonly referred to as varying electric charge theories, which have been first 
formulated by Bekenstein from a generalisation of Maxwell's equations in Ref. 
\cite{Bekenstein:1982eu}, that led to the construction of the cosmological 
varying--$e$ Bekenstein--Sandvik--Barrow--Magueijo theory of varying $\alpha$ 
\cite{Sandvik:2001rv,Barrow:2001iw,Barrow:2002db,Barrow:2002ed,Barrow:2011kr,
Barrow:2013uza}. Although, at first glance, the varying--$c$ and varying--$e$ 
theories seem to be interchangeable, each theory is characterised by its 
distinct cosmological imprints \cite{Magueijo:2002di}. Other frameworks include, 
for instance, a runaway dilaton \cite{Damour:2002mi,Damour:2002nv}, 
supersymmetric generalisation of Bekenstein's model \cite{Olive:2001vz} and a 
disformally coupled electromagnetic sector \cite{vandeBruck:2015rma}.

Several probes have  been used for the search of any space--time dependence of 
the fine--structure constant, including primarily astronomical and local 
methods. The latter ones consist of geophysical analyses of samples from the 
natural nuclear reactor in Oklo \cite{Damour:1996zw,Petrov:2005pu}, meteorites 
\cite{Olive:2003sq}, and laboratory atomic clocks 
\cite{Prestage:1995zz,Peik:2004qn,Rosenband1808}. Stringent constraints on the 
variation of the fine--structure constant have been inferred from the analysis 
of spectra from high--redshift quasar absorption systems  
\cite{Bahcall:2003rh,Murphy:2003mi,King:2012id,Agafonova:2011sp,Molaro:2013saa,
Songaila:2014fza,Evans:2014yva,Kotus:2016xxb,Murphy:2016yqp,Martins:2017yxk,
10.1093/mnras/stx179,Alves:2018mef}. Other constraints have been derived with 
the thermal Sunyaev--Zeldovich effect and X--ray measurements of galaxy clusters 
\cite{deMartino:2016tbu,Colaco:2019fvl}, strong gravitational lensing 
\cite{Liao:2015uzb,Holanda:2016msr}, and from primordial abundances of light 
nuclei produced during the era of Big Bang nucleosynthesis 
\cite{Mosquera:2013dga}. Furthermore, upcoming gravitational wave observations 
\cite{Yang:2017bkv,Fu:2019oll,Qi:2019spg} are also expected to be competitive 
with the currently available probes of the variation of the fine--structure 
constant. Moreover, the space--time dependence of fundamental constants has also 
been linked with the currently reported Hubble tension via the inferred effects 
in the ionisation history and profile of CMB anisotropies. Indeed, Ref. 
\cite{Hart:2019dxi} reported that a variation in fundamental constants, 
particularly in the effective electron mass, could play an important role in the 
alleviation of the low versus high-redshift Hubble tension.

In this work, we consider the potential variation  of the fine--structure 
parameter due to modified TG effects. These constraints are then used to limit 
the coupling parameters of literature models of $f(T)$ gravity. This is done 
using several data sets in conjunction with several literature approaches to 
parametrising the violation of the distance--duality relation which is a natural 
consequence of the violation of the EEP. This work builds on the foundations 
laid in Ref. \cite{Nunes:2016qyp} where the potential violation of the 
fine--structure constant was first studied in the context of TG. However, since 
the conformal transformations that were performed in Ref. \cite{Yang:2010ji} 
were elaborated more thoroughly in  Ref. \cite{Wright:2016ayu}, in the present 
work we   revisit the analysis of Ref. \cite{Nunes:2016qyp} and we expand its 
breadth with updated data and a deeper analysis of the potential implications.

Throughout the manuscript, Latin indices are used to refer to tangent space 
coordinates, while Greek indices refer to general manifold coordinates. The 
outline of the paper is as follows. In section~\ref{sec:f_T} we 
review TG and its extension to $f(T)$ gravity in the context of its cosmology as 
well as its potential predictions on the variation of the fine--structure 
constant. A number of $f(T)$ gravity models are constrained in 
section~\ref{sec:f_T_alpha}, in which we also discuss the cosmological 
implications of the inferred parameter constraints. Finally, the main results of 
our analyses and prospective lines of research are discussed in 
section~\ref{sec:conc}. In appendix~\ref{sec:phenom}, we also probe the general case of the 
phenomenology of a non--vanishing coupling constant in the electromagnetic 
Lagrangian which produces a violation of the distance--duality 
relation. Using literature parametrisations of this violation, we revisit and 
update the constraints on this potential violation.

\section{\label{sec:f_T}\texorpdfstring{$f(T)$}{ft} gravity and the 
fine--structure constant}

\subsection{Teleparallel Gravity}

Teleparallel gravity (TG) represents a paradigm shift in the way that gravity  is expressed not through the torsion-less connection of GR, but with the curvature-less one called the Weitzenb\"ock connection, 
$\Gamma^{\sigma}_{\mu\nu}$ \cite{Hayashi:1979qx}. In GR, curvature is calculated through the Levi-Civita connection $\mathring{\Gamma}^{\sigma}_{\mu\nu}$ \cite{nakahara2003geometry,ortin2004gravity} (recall that we use over-circles to denote quantities determined by the Levi-Civita connection). The Riemann tensor can then be used to determine a meaningful measure of curvature on a manifold, which is used in various modifications to standard gravity. Given that the Levi-Civita connection is replaced with the Weitzenb\"ock connection in TG, it follows that irrespective of the space--time under consideration, the Riemann tensor will always vanish due to the connection being curvatureless. It is for this reason that TG necessitates different measures to construct realistic models of gravity.

The dynamical objects in TG are the  tetrads $\udt{e}{a}{\mu}$, which act as a soldering agent between tangent spaces (Latin indices) and the general manifold (Greek indices) \cite{Aldrovandi:2013wha}. In this way, tetrads (and their inverses  $\dut{e}{a}{\mu}$) can be used to transform to (and from) the Minkowski metric through 
\begin{align}\label{metric_tetrad_rel}
    g_{\mu\nu}=\udt{e}{a}{\mu}\udt{e}{b}{\nu}\eta_{ab}\,,& &\eta_{ab}= 
\dut{e}{a}{\mu}\dut{e}{b}{\nu}g_{\mu\nu}\,.
\end{align}
The tetrads satisfy the  orthogonality conditions
\begin{align}
    \udt{e}{a}{\mu}\dut{e}{b}{\mu}=\delta^a_b\,,&  &\udt{e}{a}{\mu}\dut{e}{a}{\nu}=\delta^{\nu}_{\mu}\,,
\end{align}
for internal consistency. The Weitzenb\"ock connection can then be defined as \cite{Weitzenbock1923}
\begin{equation}
    \Gamma^{\sigma}_{\mu\nu}:= \dut{e}{a}{\mu}\partial_{\mu}\udt{e}{a}{\nu} + \dut{e}{a}{\sigma}\udt{\omega}{a}{b\mu}\udt{e}{b}{\nu}\,,
\end{equation}
where $\udt{\omega}{a}{b\mu}$ represents  the spin connection. This is the most general linear affine connection that is both curvatureless and satisfies the metricity condition \cite{Aldrovandi:2013wha}. Here, the spin connection appears to preserve the covariance of the resulting equations of motion 
\cite{Krssak:2015oua}. To do this, it incorporates the Local Lorentz Transformation (LLT) invariance of the theory, which implies that it can be set to zero for a particular choice of Lorentz frame \cite{Krssak:2018ywd}. 

Spin connections also appear in GR, but they are mainly hidden into the internal structure of the theory \cite{misner1973gravitation}.  Considering the full breadth of LLTs (Lorentz boosts and rotations), $\udt{\Lambda}{a}{b}$, the spin connection can be represented completely as $\udt{\omega}{a}{b\mu}=\udt{\Lambda}{a}{c}\partial_{\mu}\dut{\Lambda}{b}{c}$ \cite{Aldrovandi:2013wha}. For any particular metric tensor, there exist an infinite number of tetrads that satisfy Eq.(\ref{metric_tetrad_rel}) due to LLT invariance. Thus, it is the combination of a tetrad choice and its associated spin connection that retain the covariance of TG.

In the framework of  TG the torsion tensor is defined as \cite{Cai:2015emx}
\begin{equation}
    \udt{T}{\sigma}{\mu\nu} := 2\Gamma^{\sigma}_{[\mu\nu]}\,,
\end{equation}
where the square brackets denote the anti-symmetric  operator, and where this represents the field strength of gravitation. The torsion tensor transforms covariantly under both diffeomorphisms and LLTs. TG also relies on a couple of other tensorial quantities that help render a concise representation of the ensuing gravitational models. Firstly, the contorsion tensor turns out to be a useful quantity and is defined as the difference between the Weitzenb\"{o}ck and Levi-Civita connections, i.e.
\begin{equation}
    \udt{K}{\sigma}{\mu\nu} := \Gamma^{\sigma}_{\mu\nu} - \mathring{\Gamma}^{\sigma}_{\mu\nu} =\frac{1}{2}\left(\dudt{T}{\mu}{\sigma}{\nu} + \dudt{T}{\nu}{\sigma}{\mu} - \udt{T}{\sigma}{\mu\nu}\right)\,,
\end{equation}
which plays an important role in relating TG with Levi-Civita based theories. The second central ingredient to TG is the so-called superpotential
\begin{equation}
    \dut{S}{a}{\mu\nu}:=\frac{1}{2}\left(\udt{K}{\mu\nu}{a} - \dut{e}{a}{\nu}\udt{T}{\alpha\mu}{\alpha} + \dut{e}{a}{\mu}\udt{T}{\alpha\nu}{\alpha}\right)\,,
\end{equation}
which has been linked to the gauge current representation of  the gravitational energy-momentum tensor in TG \cite{Aldrovandi:2003pa,Koivisto:2019jra}. By contracting the torsion tensor with its superpotential produces the torsion scalar
\begin{equation}
    T:=\dut{S}{a}{\mu\nu}\udt{T}{a}{\mu\nu}\,,
\end{equation}
which is determined entirely by the Weitzenb\"{o}ck connection in  the same way that the Ricci scalar depends only on the Levi-Civita connection. By constructing the torsion scalar in this way, it turns out that the Ricci and torsion scalars are related by a total divergence term \cite{Bahamonde:2015zma,Farrugia:2016qqe}
\begin{equation}
    R=\mathring{R} + T -\frac{2}{e}\partial_{\mu}\left(e\udut{T}{\sigma}{\sigma}{\mu}\right) = 0\,,
\end{equation}
where $R$ is the Ricci scalar in terms of the Weitzenb\"{o}ck connection, which is zero, and $\mathring{R}$ is the regular Ricci scalar from GR. This implies that the Ricci and torsion scalars are equal up to a boundary term
\begin{equation}
    \mathring{R} = -T + \frac{2}{e}\partial_{\mu} \left(e\udut{T}{\sigma}{\sigma}{\mu}\right):=-T+B\,,
\end{equation}
where $e=\det\left(\udt{e}{a}{\mu}\right)=\sqrt{-g}$. This fact alone guarantees that the Ricci scalar and the torsion scalar produce the same dynamical equations. That is, we can define the TEGR action as
\begin{equation}
    \mathcal{S}_{\text{TEGR}} =  -\frac{1}{2\kappa^2}\int \mathrm{d}^4 x\; eT + \int \mathrm{d}^4 x\; e\mathcal{L}_{\text{m}}\,,
\end{equation}
where $\kappa^2=8\pi G$  and $\mathcal{L}_{\text{m}}$ is the matter Lagrangian. While both actions lead to the same dynamical equations, they differ in terms of their Lagrangian in that the TG formulation decouples the second-order derivative contributions to the field equations, and the fourth-order derivative contribution which appears as a boundary quantity. This is not relevant for comparing GR and TEGR, but becomes an active agent when modifications to gravity are considered.

Using the same reasoning that led to $f(\mathring{R})$ gravity \cite{DeFelice:2010aj,Capozziello:2011et}, the Lagrangian of TEGR can be generalised to $f(T)$ gravity  \cite{Ferraro:2006jd,Ferraro:2008ey,Bengochea:2008gz,Linder:2010py,Chen:2010va}, giving
\begin{equation}
    \mathcal{S}_{\text{TEGR}} =  \frac{1}{2\kappa^2}\int \mathrm{d}^4 x\; ef(T) + \int \mathrm{d}^4 x\; e\mathcal{L}_{\text{m}}\,.
\end{equation}
This  produces   second-order equations, which is only possible since 
the Lovelock theorem is much weaker in TG 
\cite{Lovelock:1971yv,Gonzalez:2015sha,Bahamonde:2019shr}.
Note that TG and thus GR, are reproduced if $f(T)=-T+\Lambda$.
$f(T)$ gravity also 
shares a number of properties with GR such as the same gravitational wave polarisation structure 
\cite{Farrugia:2018gyz,Cai:2018rzd,Abedi:2017jqx,Chen:2019ftv}, and being 
Gauss-Ostrogradsky ghost free (since it remains second-order) 
\cite{Krssak:2018ywd,ortin2004gravity}. Finally, by performing variation of 
the $f(T)$ action with respect to the tetrads, we arrive at the following field 
equations
\begin{align}\label{ft_FEs}
    e^{-1} &\partial_{\nu}\left(e\dut{e}{a}{\rho}\dut{S}{\rho}{\mu\nu}\right)f_T 
-  \dut{e}{a}{\lambda} \udt{T}{\rho}{\nu\lambda}\dut{S}{\rho}{\nu\mu} f_T + 
\frac{1}{4}\dut{e}{a}{\mu}f(T) \nonumber\\
    & + \dut{e}{a}{\rho}\dut{S}{\rho}{\mu\nu}\partial_{\nu}\left(T\right)f_{TT}  + \dut{e}{b}{\lambda}\udt{\omega}{b}{a\nu}\dut{S}{\lambda}{\nu\mu}f_T =  
\kappa^2 \dut{e}{a}{\rho} \dut{\Theta}{\rho}{\mu}\,,
\end{align}
where subscripts denote derivatives, and $\dut{\Theta}{\rho}{\nu}$ is the 
regular energy-momentum tensor.

\subsection{\label{sec:fT_cosmology}\texorpdfstring{$f(T)$}{ft} cosmology}

We investigate the cosmology   of $f(T)$ gravity through a flat homogeneous and 
isotropic metric. We consider a tetrad choice of the form
\begin{equation}
    \udt{e}{a}{\mu}=\text{diag}\left(1,\,a(t),\,a(t),\,a(t)\right)\,,
\end{equation}
where $a(t)$ is the scale factor, and which allows us to set the spin 
connection to zero, i.e. $\udt{\omega}{a}{b\mu}=0$ 
\cite{Krssak:2015oua,Tamanini:2012hg}. Through Eq. (\ref{metric_tetrad_rel}), the 
flat FLRW metric is reproduced
 \begin{equation}\label{FLRW_metric}
     \mathrm{d}s^2=-\mathrm{d}t^2+a^2(t) \left(\mathrm{d}x^2+\mathrm{d}y^2+\mathrm{d}z^2\right)\,.
 \end{equation}
Straightforwardly, we can calculate the torsion scalar to be
\begin{equation}\label{Tor_sca_flrw}
    T=6H^2\,,
\end{equation}
and the boundary term to be $B=6\left(3H^2+\dot{H}\right)$, which reproduces the 
well-known Ricci scalar for this  metric, i.e. 
$\mathring{R}=-T+B=6\left(\dot{H}+2H^2\right)$ (note that we   use the  
standard convention for the metric signature \cite{Kofinas:2014owa}, instead of 
the one used in Refs. \cite{Bengochea:2008gz,Linder:2010py,Cai:2015emx}, which leads 
to a sign difference in $T$). By evaluating the field 
equations in Eq. (\ref{ft_FEs}), the resulting Friedmann equations are
\begin{align}
    H^2 &= \frac{\kappa^2}{3}\left(\rho +  
\rho_{\text{eff}}\right)\,,\label{Friedmann_1}\\
    \dot{H} &= -\frac{\kappa^2}{2} \left(\rho + \rho_{\text{eff}} + p + p_{\text{eff}}\right)\label{Friedmann_2}\,,
\end{align}
where $\rho$ and $p$ represent the energy  density and pressure of the matter 
content respectively, while $f(T)$ gives rise to  an effective fluid 
with components
\begin{align}
    \rho_{\text{eff}} &:= \frac{1}{2\kappa^2}\left(T-f+2Tf_T\right)\,,\\
    p_{\text{eff}} &:= -\frac{1}{2\kappa^2}\left[4\dot{H}\left(1+f_T+2Tf_{TT}\right)\right] - \rho_{\text{eff}}\,,
\end{align}
which also satisfies the conservation equation
\begin{equation}
    \dot{\rho}_{\text{eff}} + 3H\left(\rho_{\text{eff}}+p_{\text{eff}}\right) = 0\,.
\end{equation}
In this way, we can define an equation of state (EoS) of the effective fluid as 
\cite{Farrugia:2016qqe}
\begin{align}
    \omega_{\text{eff}} &:= \frac{p_{\text{eff}}}{\rho_{\text{eff}}}\nonumber\\
    &= -1 + \frac{4\dot{H}\left(1+f_T+2Tf_{TT}\right)}{T-f+2Tf_T}\nonumber\\
    &= -1 
+\left(1+\omega_m\right)\frac{\left(f-2Tf_T\right)\left(1+f_T+2Tf_{TT}\right)}{
\left(f_T+2Tf_{TT}\right)\left(T-f+2Tf_T\right)}\,,
\end{align}
where the last line is a result of the Friedmann equations in 
Eqs. (\ref{Friedmann_1},\ref{Friedmann_2}), and $\omega_m$ is the EoS of matter. 
Notice that we recover the $\Lambda$CDM scenario ($\omega_{\text{eff}}=-1$) for 
  $f(T)=-T+\Lambda$. Finally, by considering scalar perturbations on the 
flat FLRW of Eq. (\ref{FLRW_metric}) together with matter perturbations, an 
effective Newton's constant can be defined as in Refs. 
\cite{Zheng:2010am,Wu:2012hs,Nunes:2018xbm} such that $G_{\text{eff}} = 
 \frac{G_{N}}{|f_T|}\,,$ where $G_{N}^{}$ is Newton's constant.

\subsection{The fine--structure constant in Teleparallel Gravity}

The fine--structure constant and the luminosity distance are derived from  the 
electromagnetic action \cite{PhysRevD.90.023017,Hees:2014lfa} which is 
conformally invariant \cite{peebles:1993}. Conformal transformations are 
important because for many theories of modified gravity, they can be used to 
transform between the Jordan and Einstein frames 
\cite{PhysRevD.39.3159,Carloni:2009gp,Capozziello:2011et}, where the extra 
degrees of freedom of a theory may appear as   scalar fields. There exists a 
number of theories of gravity in which conformal transformations do not lead to 
the Einstein frame. This implies that the Einstein frame would be a result of 
another type of transformation which may produce a coupling with the 
electromagnetic Lagrangian 
\cite{vandeBruck:2015rma,Belgacem:2017ihm,Wetterich:2003jt,Damour:2002nv,
Hees:2014yba,Lee:2004vm,Copeland:2003cv}. This also occurs when the low-energy 
limit of quantum gravity theories are considered 
\cite{Hees:2014lfa,Brax:2010uq,Damour:2002mi,Hees:2014lfa,PhysRevD.90.023017},
which may appear as heavy fermions  for instance. In either case, the result is 
the introduction of a new degree of freedom, $\phi$, that arises from the 
transformation
\begin{equation}
    \udt{e}{a}{\mu} \rightarrow \udt{\tilde{e}}{a}{\mu}\,,
\end{equation}
where $\udt{\tilde{e}}{a}{\mu}$ represents the Einstein frame tetrad.  This then 
induces an electromagnetic coupling which takes on the form
\begin{equation}\label{emag_action}
    S_{\text{EM}} = -\frac{1}{4} \int \mathrm{d}^4 x\, e B_F\left(\phi\right) F_{\mu\nu}F^{\mu\nu}\,,
\end{equation}
where $F_{\mu\nu}=A_{\nu,\mu} -  A_{\mu,\nu}$ is the standard Faraday tensor, 
and $B_F\left(\phi\right)$ represents the nonvanishing $\phi-$coupling. The 
consequence of this induced coupling is that the fine--structure constant and the 
luminosity distance will be altered comparing to   GR   
\cite{misner1973gravitation,PhysRevD.90.023017}. As in Refs.  
\cite{Brax:2012gr,Olive:2001vz,Copeland:2003cv,Nunes:2016plz}, this can be 
expanded about $\phi(t=t_0)$, which is suitably small, to give
\begin{equation}\label{B_F_Var_order}
    B_F\left(\phi\right) \simeq 1+\beta_F \frac{\phi}{M_{\text{pl}}}\,,
\end{equation}
where $\beta_F=\mathcal{O}(1)$ is a  constant, and $M_{\text{pl}}=1/\kappa^2$ is 
the Planck mass ($\beta_F\phi<<M_{\text{pl}}$).

Given an initially uncoupled Jordan--frame electromagnetic action, the 
fine--structure constant turns out to be given by \cite{Olive:2001vz}
\begin{equation}
    \alpha_E (\phi) = \frac{\alpha_J(\phi)}{B_F(\phi)}\,,
\end{equation}
where $\alpha_E$ and $\alpha_J$ are the fine--structure constants  in the 
Einstein and Jordan frames respectively. To relate a change in the 
fine--structure constant between these frames with the electromagnetic coupling 
term in Eq. (\ref{emag_action}), consider the fractional change 
\cite{Nunes:2016plz,Hees:2014lfa}
\begin{equation}
    \frac{\Delta\alpha}{\alpha} := \frac{\alpha_E-\alpha_J}{\alpha_J} =  
\frac{1}{B_F\left(\phi\right)} - 1\,,
\end{equation}
which depends on redshift (or cosmic time). Since $B_F(z=0) := B_{F_0} \neq 1$,  
we need to rescale this relation so that $\Delta \alpha=0$ at present time 
($z=0$). This can be conveniently done by taking $B_F\left(\phi\right) 
\rightarrow B_F\left(\phi\right)/B_{F_0}$ which is a result of the Maxwell 
tensor transformation $F_{\mu\nu} \rightarrow \sqrt{B_{F_0}} F_{\mu\nu}$. Hence, 
the fractional change in the fine--structure constant now emerges as
\begin{equation}
    \frac{\Delta \alpha}{\alpha} = \frac{B_{F_0}}{B_F\left(\phi\right)} - 1\,.
\end{equation}

In $f(T)$ gravity, the form  of this fine--structure constant dependence can be 
obtained by a conformal transformation of the tetrad where
\begin{align}
    \udt{\tilde{e}}{a}{\mu} = \Omega\,\udt{e}{a}{\mu}\,,& &\dut{\tilde{e}}{a}{\mu} = \Omega^{-1}\,\dut{e}{a}{\mu}\,,
\end{align}
which results in the regular conformal transformation 
$\tilde{g}_{\mu\nu}=\Omega^2 g_{\mu\nu}$, as expanded upon in Ref.  
\cite{Wright:2016ayu}, where $\Omega^2 = -f_T=|f_T|$ (note that since in our 
conventions $T>0$ and $f_T<0$, 
we have replaced $-f_T$ 
by $|f_T|$). It is well-known that  $f(T)$ 
gravity cannot be written in the Einstein frame through conformal 
transformations, which implies that it will induce a dependence in its 
associated fine--structure constant characterised by Eq. (\ref{B_F_Var_order}) 
\cite{Brax:2012gr}. In fact, this produces an extra 
$2\Omega^{-6}\tilde{\partial}^{\mu}\left(\Omega^2\right)\udt{\tilde{T}}{\nu}{
\nu\mu}$ term which cannot be removed. The remainder of the scalar field becomes 
a phantom field with the choice of $\phi=\sqrt{3}\ln f_T$ \cite{Wright:2016ayu}, 
which is partially favored by recent \textit{Planck} data \cite{Aghanim:2018eyx} (this 
form of the scalar field is a correction to the one used in Ref.  
\cite{Nunes:2016plz} due to a typo in Ref. \cite{Yang:2010ji}).

Recent constraints on observationally viable  models of $f(T)$ gravity 
\cite{Nesseris:2013jea,Farrugia:2016xcw,Iorio:2015rla,Wu:2010mn,
Capozziello:2017uam,Nunes:2018xbm,Nunes:2016qyp} suggest that the model 
Lagrangian would take the form of TEGR plus small corrections. Given that 
$\Omega^2 =  |f_T|$, this would imply that the 
$\tilde{\partial}^{\mu}\left(\Omega^2\right)$ would be very small rendering the 
additional term negligible. We will revisit this reasoning against the results 
of the analysis. With this approximation to the Einstein frame, the variation of 
the fine--structure constant takes the form
\begin{equation}\label{f_T_fine_struc_const}
    \frac{\Delta \alpha}{\alpha} = \frac{M_{\text{pl}} + \sqrt{3}\beta_F 
 [\ln |f_T(T_0)|]}{M_{\text{pl}} + \sqrt{3}\beta_F [\ln |f_T|]} - 1\,,
\end{equation}
which vanishes for the $\Lambda$CDM case of $f(T)=-T+\Lambda$, as  expected. 
Eq. (\ref{f_T_fine_struc_const}) embodies the redshift dependence of the 
fine--structure constant in TG, since the torsion scalar depends 
on redshift in accordance with Eq. (\ref{Tor_sca_flrw}). Another consequence of a 
nonvanishing scalar field coupling to the electromagnetic action is that the 
luminosity distance will be altered \cite{Hees:2014lfa}. By considering the 
standard derivation of luminosity distance \cite{misner1973gravitation} with 
this new action, Ref. \cite{PhysRevD.90.023017} shows that this leads to
\begin{align}\label{eq:dL_f_T}
    d_L &= c\left(1+z\right)\sqrt{\frac{B_{F_0}}{B_F}} \int_0^z  
\frac{\mathrm{d}z}{H(z)}\nonumber\\
    &=c\left(1+z\right) \sqrt{\frac{M_{\text{pl}} +  \sqrt{3}\beta_F [\ln 
|f_T(T_0)|]}{M_{\text{pl}} + \sqrt{3}\beta_F [\ln 
|f_T|]}} \int_0^z 
\frac{\mathrm{d}z}{H(z)}\,,
\end{align}
as the luminosity distance for $f(T)$ gravity, which limits to the GR formula for $B_F=1$.

\begin{table*}[t]
    \centering
    \setlength\extrarowheight{4pt}
    \begin{tabular}{ c c c c }
        \hline
        \hline
        \multicolumn{4}{c}{$f_1^{}(T)$ Model}\\
        Parameter & SN + CC + $H_0^R$ & SN + CC + KVNO & SN + CC + KVNO + $H_0^R$ \\ 
		\hline
		$b$ & $-0.16^{+0.24}_{-0.49}$ & ~~$0.003^{+0.053}_{-0.059}$ & $-0.001^{+0.050}_{-0.048}$ \\ 
		$\Omega^\mathrm{m}_0$ & ~~$0.281^{+0.036}_{-0.035}$ & ~~$0.300^{+0.026}_{-0.024}$ & ~~$0.283^{+0.023}_{-0.021}$ \\ 
		$H_0^{}$ & ~$72.8^{+1.4}_{-1.3}$ & ~$68.9^{+2.0}_{-1.9}$ & ~$72.2^{+1.2}_{-1.2}$ \\ 
		$\beta_F$ & ~~$0.28^{+0.32}_{-0.32}$ & $-0.003^{+0.063}_{-0.056}$ & $-0.003^{+0.074}_{-0.067}$~ \\[.5em]
		\hline
		\hline
        \multicolumn{4}{c}{$f_2^{}(T)$ Model}\\
        Parameter & SN + CC + $H_0^R$ & SN + CC + KVNO & SN + CC + KVNO + $H_0^R$ \\ 
		\hline
		$1/p$ & ~~$0.093^{+0.171}_{-0.079}$ & $\left( 10.8^{+35.9}_{-4.9} \right) \times 10^{-3}$ & $\left( 41.7^{+9.3}_{-30.8} \right) \times 10^{-3}$ \\ 
		$\Omega^\mathrm{m}_0$ & ~~$0.279^{+0.025}_{-0.031}$ & ~~$0.300^{+0.021}_{-0.020}$ & ~~$0.283^{+0.020}_{-0.019}$ \\ 
		$H_0^{}/\,\mathrm{km}\,\mathrm{s}^{-1}\mathrm{Mpc}^{-1}$ & ~$72.2^{+1.3}_{-1.2}$ & ~$69.0^{+1.8}_{-1.9}$ & ~$72.2^{+1.3}_{-1.2}$ \\ 
		$\beta_F$ & $-0.10^{+0.49}_{-0.56}$ & $-0.01^{+0.45}_{-0.75}$ & $-0.07^{+0.58}_{-0.55}$ \\[.5em]
		\hline
		\hline
		\multicolumn{4}{c}{$f_3^{}(T)$ Model}\\
        Parameter & SN + CC + $H_0^R$ & SN + CC + KVNO & SN + CC + KVNO + $H_0^R$ \\ 
		\hline
		$1/q$ & ~~$0.065^{+0.088}_{-0.045}$ & $\left( 15.7^{+28.7}_{-9.5} \right) \times 10^{-3}$ & ~$0.029^{+0.018}_{-0.020}$ \\ 
		$\Omega^\mathrm{m}_0$ & ~~$0.279^{+0.021}_{-0.020}$ & ~~$0.302^{+0.022}_{-0.023}$ & ~~$0.283^{+0.019}_{-0.019}$~ \\ 
		$H_0^{}/\,\mathrm{km}\,\mathrm{s}^{-1}\mathrm{Mpc}^{-1}$ & ~$72.2^{+1.3}_{-1.2}$ & ~~$69.0^{+2.0}_{-1.9}$~ & ~$72.2^{+1.2}_{-1.3}$~ \\ 
		$\beta_F$ & $-0.22^{+0.76}_{-0.40}$ & $-0.05^{+0.61}_{-0.53}$ & ~~$0.00^{+0.61}_{-0.52}$~ \\[.5em]
		\hline
		\hline
		\multicolumn{4}{c}{$f_4^{}(T)$ Model}\\
        Parameter & SN + CC + $H_0^R$ & SN + CC + KVNO & SN + CC + KVNO + $H_0^R$ \\ 
		\hline
		$m$ & $1.07^{+6.04}_{-0.50}$ & $1.16^{+5.51}_{-0.67}$ & $1.16^{+5.53}_{-0.65}$ \\ 
		$\Omega^\mathrm{m}_0$ & $0.264^{+0.029}_{-0.033}$ & $0.301^{+0.020}_{-0.021}$ & $0.282^{+0.019}_{-0.017}$ \\ 
		$H_0^{}/\,\mathrm{km}\,\mathrm{s}^{-1}\mathrm{Mpc}^{-1}$ & $72.5^{+1.4}_{-1.1}$~ & $69.0^{+1.9}_{-1.8}$~ & $72.2^{+1.1}_{-1.2}$~ \\ 
		$\beta_F$ & $-0.25^{+0.37}_{-0.32}$~~ & $\left( -1.6^{+2.6}_{-2.3} \right) \times 10^{-6}$ & $\left( -1.5^{+2.5}_{-2.4} \right) \times 10^{-6}$ \\[.5em]
		\hline
		\hline
		\multicolumn{4}{c}{$f_5^{}(T)$ Model}\\
        Parameter & SN + CC + $H_0^R$ & SN + CC + KVNO & SN + CC + KVNO + $H_0^R$ \\ 
		\hline
		$n$ & ~$1.49^{+0.25}_{-0.15}$~ & $1.934^{+0.055}_{-0.331}$ & $1.941^{+0.047}_{-0.306}$ \\ 
		$\Omega^\mathrm{m}_0$ & $0.298^{+0.076}_{-0.091}$ & $0.291^{+0.051}_{-0.155}$ & $0.147^{+0.156}_{-0.028}$ \\ 
		$H_0^{}/\,\mathrm{km}\,\mathrm{s}^{-1}\mathrm{Mpc}^{-1}$ & $72.8^{+1.4}_{-1.4}$~ & $68.5^{+2.2}_{-1.9}$~ & $72.2^{+1.2}_{-1.3}$~ \\ 
		$\beta_F$ & ~$0.039^{+0.043}_{-0.037}$~ & $\left( -6.8^{+0.1}_{-7.4} \right) \times 10^{-4}$ & $\left( 3.8^{+10.5}_{-9.1} \right) \times 10^{-7}$ \\[.5em] 
		\hline
		\hline
    \end{tabular}
    \caption{The mean value and the corresponding 68\% limits of the model parameters of the five $f_{i}(T)$ models ($i\in\{1,2,3,4,5\}$), as described in section \ref{sec:f_T_constraints}.}
    \label{tab:f_models}
\end{table*}

\section{\label{sec:f_T_alpha}Probing \texorpdfstring{$f(T)$}{ft}  gravity by 
its induced variation in \texorpdfstring{$\alpha$}{alpha}}

In this section we present   the inferred constraints on five 
distinct $f(T)$ gravitational models by adopting a Bayesian approach for each model under consideration. This was 
implemented in the Markov chain Monte Carlo (MCMC) Ensemble sampler 
\texttt{emcee} \cite{ForemanMackey:2012ig}. We then analysed our 
chains by the publicly available package \texttt{ChainConsumer} 
\cite{Hinton2016}.

We consider flat priors for 
all the varied $f(T)$ model parameters 
$\Theta=\{\chi,\,\Omega_0^\mathrm{m},\,H_0^{},\,\beta_F^{}\}$, where $\chi$ is 
the specific model parameter characterising each particular model which will be 
discussed in the next section, $\Omega^\mathrm{m}_0$ is the dimensionless energy 
density of pressureless matter today, $H_0^{}$ denotes the Hubble's 
constant, and $\beta_F^{}$ is the electromagnetic coupling constant defined in 
Eq. (\ref{B_F_Var_order}).

We have independently and jointly considered the measurements  of 
$\Delta\alpha/\alpha$ from the archival astrophysical data measurements from 
quasar absorption lines observed at the Keck (K) observatory 
\cite{Murphy:2003mi} and with the VLT (V) \cite{King:2012id}, along with a set 
of 21 dedicated new measurements (N) 
\cite{Agafonova:2011sp,Molaro:2013saa,Songaila:2014fza,Evans:2014yva,
Kotus:2016xxb,Murphy:2016yqp,Martins:2017yxk,
10.1093/mnras/stx179,Alves:2018mef}, and the constraint from the Oklo (O) 
natural nuclear reactor at an effective redshift of $z=0.14$ 
\cite{Petrov:2005pu}. We remark that the measurements contained in the N data 
set were reported from the ESO Ultraviolet and Visual Echelle Spectrograph 
(UVES) Large Program which was specifically developed for such measurements. In what follows, we 
denote the joint data sets of: $\mathrm{N}+\mathrm{O}$ by NO, 
$\mathrm{K}+\mathrm{V}$ by KV, $\mathrm{K}+\mathrm{V}+\mathrm{N}$ by KVN, and 
$\mathrm{K}+\mathrm{V}+\mathrm{N}+\mathrm{O}$ by KVNO. 

Additionally, we will be 
making use of the Supernovae Type Ia (SN) Pantheon Sample 
\cite{Scolnic:2017caz}, and a cosmic chronometers (CC) data set 
\cite{Moresco:2016mzx,Moresco:2012jh,Simon:2004tf,Stern:2009ep,Zhang:2012mp,
Moresco:2015cya} composed of Hubble parameter measurements which are determined 
from the differential age of old and passive evolving galaxies 
\cite{Jimenez:2001gg}. We further adopt a prior 
likelihood $(H_0^R)$ on the Hubble constant of 
$H_0^{}=74.03\pm1.42\,\mathrm{km}\,\mathrm{s}^{-1}\mathrm{Mpc}^{-1}$ 
\cite{Riess:2019cxk}, in order to check for any model parameter dependencies on the value of the Hubble constant.

\subsection{\label{sec:f_T_constraints}Current constraints}

We will be considering five distinct $f(T)$ models, which have been extensively studied in 
the literature and found to be cosmologically viable (see, for instance, Ref. \cite{Nesseris:2013jea}). The inferred mean values and 68\% limits are reported in Table \ref{tab:f_models}, and the obtained results are discussed in the below sections.

\subsubsection{The $f_1(T)$ model}

We consider the power--law model \cite{Bengochea:2008gz} as the first $f(T)$ model, specified by

\begin{equation}
\label{fTmodel1}
    f_1(T)=-T+\alpha_1\,T^b\,,
\end{equation}
where $\alpha_1$ and $b$ are constant model parameters, such that
\begin{equation}
    \alpha_1=\left(6H_0^2\right)^{1-b}\frac{1-\Omega^\mathrm{m}_0}{2b-1}\,,
\end{equation}
which follows from Eq. (\ref{Friedmann_1}). Clearly,  we recover the 
$\Lambda$CDM model when $b=0$, while this $f(T)$ model can mimic the 
Dvali--Gabadadze--Porrati (DGP) model \cite{Dvali:2000hr} when $b=1/2$.

\begin{figure}[t]
\begin{center}
    \includegraphics[width=0.88\columnwidth]{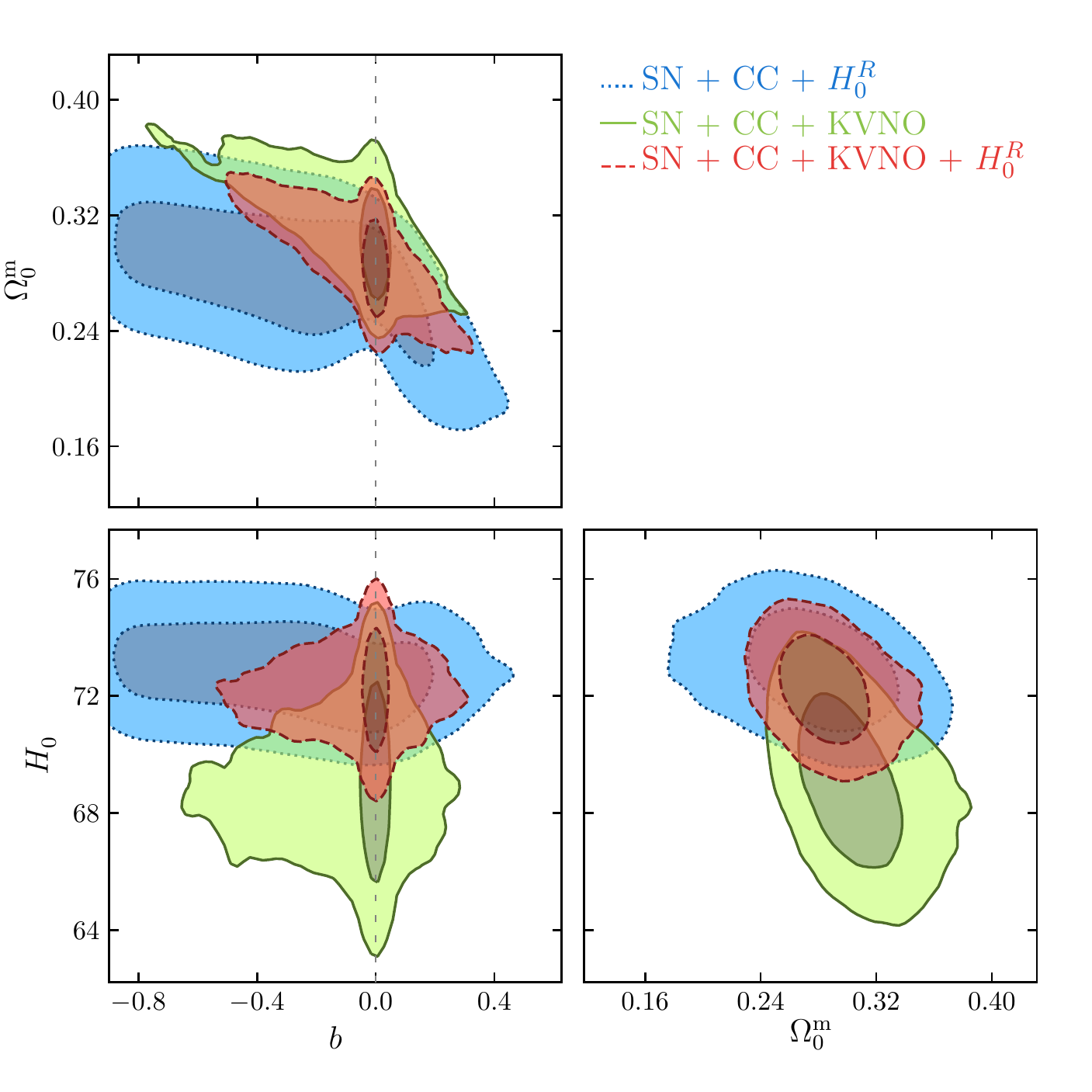}
    \caption{\label{fig:f1CDM_posteriors}{Marginalised two--dimensional likelihood constraints on the parameters of the $f_1^{}(T)$ model of Eq. (\ref{fTmodel1}).}}
\end{center}
\end{figure}

The inferred model parameter constraints are reported in  the top panel of Table 
\ref{tab:f_models}, and in Fig. \ref{fig:f1CDM_posteriors} we illustrate the 
marginalised two--dimensional likelihood constraints. We should remark that even 
without the varying fine--structure constant observational probes, we were still 
able to impose a robust constraint on $\beta_F^{}$ via the SN likelihood. As 
expected, the constraints on $\beta_F^{}$ improve significantly when we further 
make use of the KVNO data set, from which we find that $\beta_F^{}$ is 
compatible with zero. Consequently, there is a negligible deviation from the 
$f(T)$ distance--duality relation in this model.

The derived constraint on  $b=-0.16^{+0.24}_{-0.49}$ from the SN + CC + $H_0^R$ 
joint data set is consistent with the findings in previous studies 
\cite{Nesseris:2013jea,Qi:2017xzl,Xu:2018npu,Anagnostopoulos:2019miu}, which 
however did not consider a varying $\beta_F^{}$. As illustrated in Fig. 
\ref{fig:f1CDM_posteriors}, the SN + CC + KVNO and SN + CC + KVNO + $H_0^R$ 
joint data sets also improve the constraints on the model parameter $b$, which 
we find to be consistent with zero. Consequently, our constraints are tighter 
than the ones reported in a similar analysis of Ref. \cite{Nunes:2016plz}. 
Indeed, all results show that the $f_1(T)$ model is in agreement with the 
$\Lambda$CDM model at the 1$\sigma$ level, in line with Refs. 
\cite{Nesseris:2013jea,Nunes:2016plz,Qi:2017xzl,Xu:2018npu,
Anagnostopoulos:2019miu}, and different from Ref. \cite{Nunes:2016qyp}. 

\begin{figure}[t]
\begin{center}
    \includegraphics[width=0.88\columnwidth]{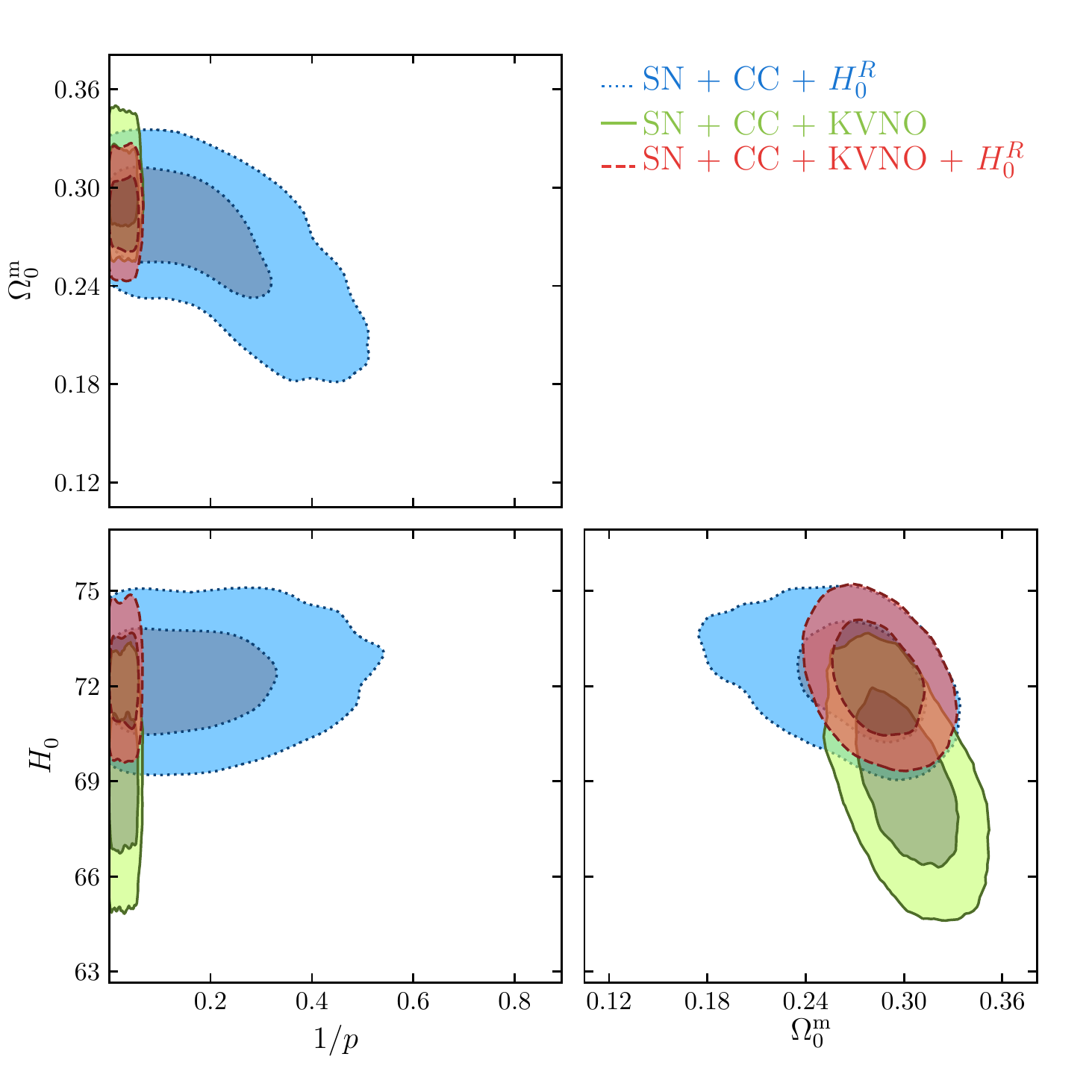}
    \caption{\label{fig:f2CDM_posteriors}{Marginalised two--dimensional likelihood constraints on the parameters of the $f_2^{}(T)$ model specified in Eq. (\ref{fTmodel2}).}}
\end{center}
\end{figure}

\subsubsection{The $f_2(T)$ model}

The second $f(T)$ model is the square--root--exponential model 
\cite{Linder:2010py}  given by
\begin{equation}
\label{fTmodel2}
    f_2(T)=-T+\alpha_2\, T_0\left(\,1-e^{-p\sqrt{T/T_0}}\,\right)\,,
\end{equation}
with model parameters $\alpha_2$ and $p$, and we recall that 
$T_0=T(z=0)=6H_0^2$ denotes the current  torsion scalar. From the Friedmann Eq. 
(\ref{Friedmann_1})  we find that
\begin{equation}
    \alpha_2=-\frac{1-\Omega^\mathrm{m}_0}{1-(1+p)e^{-p}}\,.
\end{equation}
Thus, for $p\rightarrow+\infty$, the $f_2(T)$ model reduces to  the concordance 
$\Lambda$CDM model. In our analysis we therefore vary the parameter $1/p$, for 
which we recover the $\Lambda$CDM model when $1/p\rightarrow0^+$.

The inferred constraints on $1/p$ from the joint data sets of SN + CC + $H_0^R$ 
and SN + CC + KVNO + $H_0^R$ are consistent with zero at around 1$\sigma$, 
while the $f_2(T)$ model is found to be in agreement with the $\Lambda$CDM 
model at around 2$\sigma$ when the SN + CC + KVNO data set is adopted. This 
observation is in line with other studies, such as Refs. 
\cite{Nesseris:2013jea,Qi:2017xzl,
Nunes:2016plz,Nunes:2016qyp,Xu:2018npu,Anagnostopoulos:2019miu}. We depict the 
marginalised confidence contours in Fig. \ref{fig:f2CDM_posteriors} and we list 
all the derived constraints in the second panel of Table \ref{tab:f_models}.

This $f(T)$ model is  also found to be consistent with the distance--duality 
relation, although the KVNO data set did not significantly improve the 
constraints on $\beta_F^{}$, which were always found to be in agreement with 
zero. However, the variation of the fine--structure constant relationship of 
Eq. (\ref{f_T_fine_struc_const}) led to tighter constraints on $1/p$ than those 
reported in Ref. \cite{Nunes:2016plz}.

\begin{figure}[t]
\begin{center}
    \includegraphics[width=0.88\columnwidth]{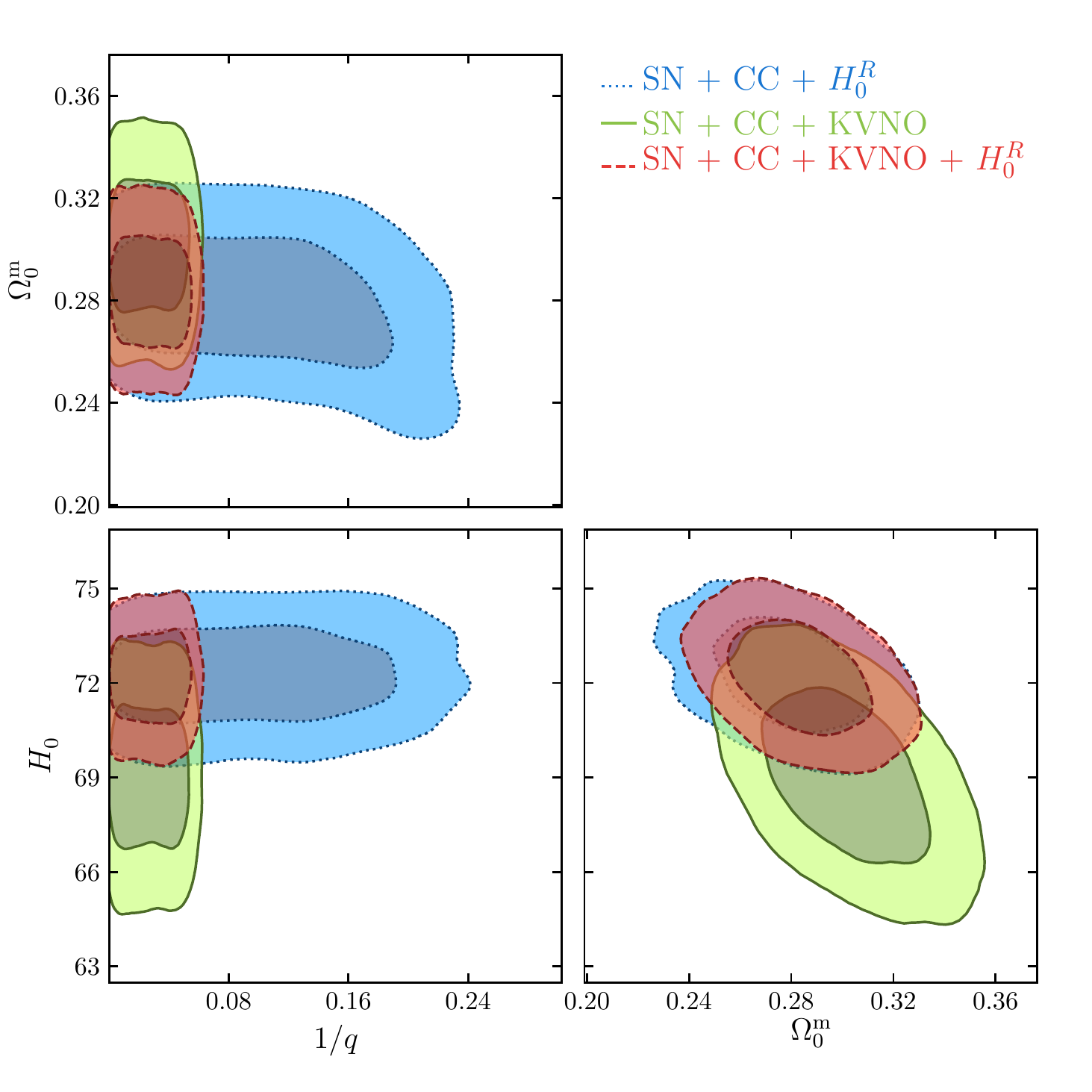}
    \caption{\label{fig:f3CDM_posteriors}{Marginalised two--dimensional likelihood constraints on  the   parameters of the $f_3^{}(T)$ model of Eq. (\ref{fTmodel3}).}}
\end{center}
\end{figure}

\subsubsection{The $f_3(T)$ model}

A similar model to the $f_2(T)$ model  is the exponential model 
\cite{Linder:2010py}, which is also motivated by $f(R)$ gravity 
\cite{Linder:2009jz}, and is given by
\begin{equation}
\label{fTmodel3}
    f_3(T)=-T+\alpha_3\, T_0\left(1-e^{-qT/T_0}\right)\,,
\end{equation}
with 
\begin{equation}
    \alpha_3=\frac{1-\Omega^\mathrm{m}_0}{-1+(1+2q)e^{-q}}\,,
\end{equation}
and $q$ is the remaining model parameter. Again, we observe that the  
$\Lambda$CDM model is recovered when $q\rightarrow+\infty$, or equivalently 
$1/q\rightarrow0^+$. For convenience, we will be considering $1/q$ as our free 
parameter.

We report the derived parameter constraints in the third panel of Table 
\ref{tab:f_models} and we depict the marginalised confidence contours in Fig. 
\ref{fig:f3CDM_posteriors}. Similar to the previous exponential $f(T)$ model, 
the model parameter $1/q$ is found to be consistent with the $\Lambda$CDM limit 
at around 1$\sigma$. This is compatible with the results of Refs. 
\cite{Nesseris:2013jea,Nunes:2016qyp,Xu:2018npu,Anagnostopoulos:2019miu}, 
although in these studies the possible variation of the fine--structure 
constant has not been taken into account.

Moreover, this exponential  $f(T)$ model is characterised by a null variation 
in the fine--structure constant, since $\beta_F^{}$ is always found to be 
consistent with zero. However, we note that the KVNO data set did not 
significantly ameliorate the constraints inferred by the SN + CC + $H_0^R$ 
joint data set.

\begin{figure}[t]
\begin{center}
    \includegraphics[width=0.88\columnwidth]{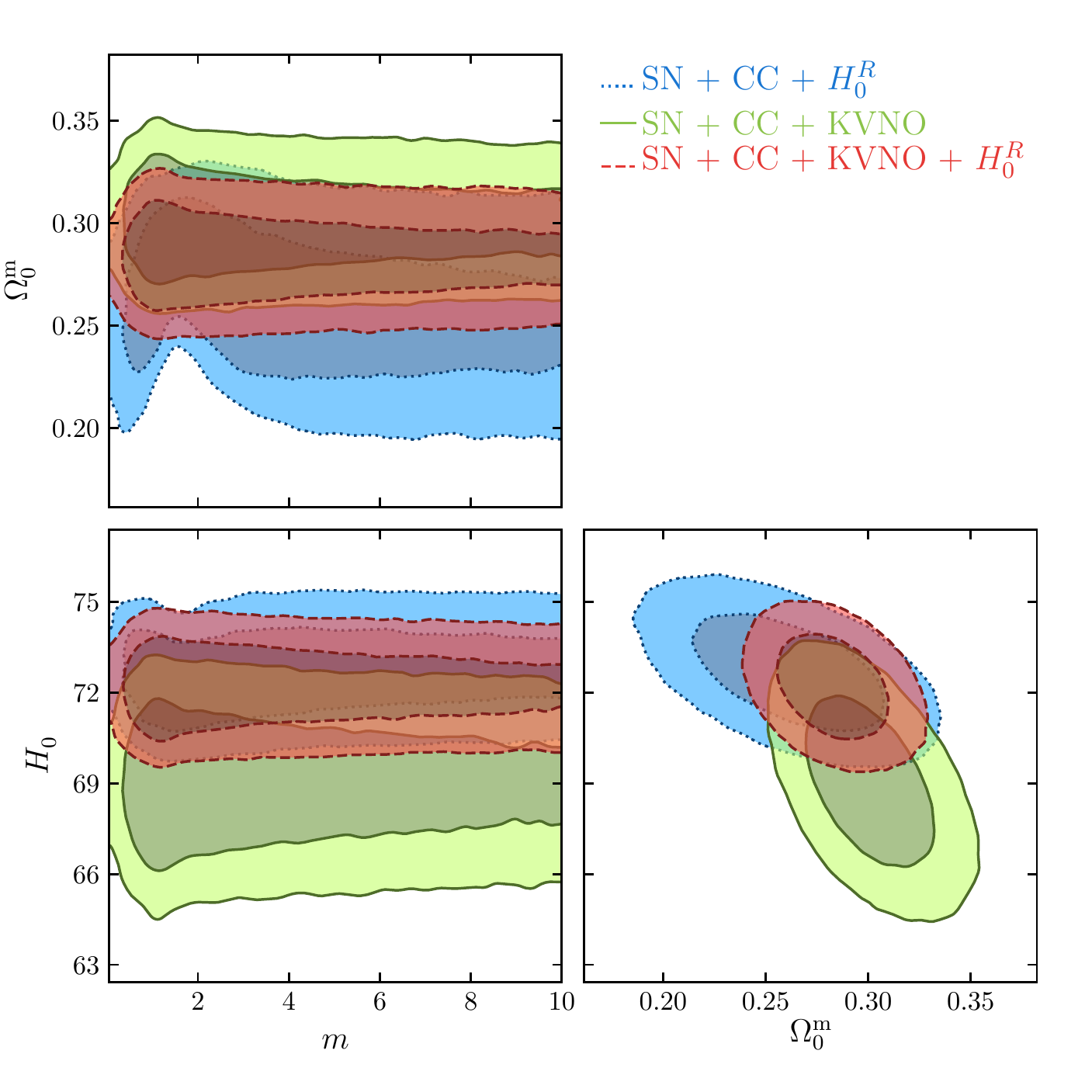}
    \caption{\label{fig:f4CDM_posteriors}{Marginalised two--dimensional likelihood constraints on  the   parameters of the $f_4^{}(T)$ model specified by Eq. (\ref{fTmodel4}).}}
\end{center}
\end{figure}

\subsubsection{The $f_4(T)$ model}

The next model which will be considered in our analysis  is the logarithmic 
model \cite{Bamba:2010wb}, given by
\begin{equation}
    f_4(T)=-T+\alpha_4\, 
T_0\sqrt{\frac{T}{mT_0}}\ln\left(\frac{mT_0}{T}\right)\,,
\label{fTmodel4}
\end{equation}
such that
\begin{equation}
    \alpha_4=-\frac{\left(1-\Omega^\mathrm{m}_0\right)\sqrt{m}}{2}\,,
\end{equation}
and $m$ is the model parameter which will  be varied in our MCMC analysis. We 
should remark that unlike the previously considered $f(T)$ models, this model 
cannot reduce to the concordance model of cosmology for any chosen value of $m$.

Interestingly enough, the background evolution of this model coincides with 
that  of the spatially flat self--accelerating branch of the DGP braneworld 
model \cite{Dvali:2000hr,Deffayet:2000uy}. However, the evolution of 
cosmological perturbations differ from one model to another. For instance, the 
functional forms of $G_\mathrm{eff}/G_N$ are not identical. Obviously, the 
well--known significant inconsistencies of the spatially flat 
self--accelerating DGP model with cosmological data (see e.g. Refs. 
\cite{Maartens:2006yt,Xu:2016grp,Lombriser:2009xg}  and references therein) 
will be inherited by the $f_4(T)$ model, and we therefore expect that this 
model will not be viable.

Moreover, the resulting Friedmann equation is independent from the model  
parameters $\alpha_4$ and $m$, in contrast with all the other $f(T)$ models 
considered in this section. Consequently, no constraints can be placed on the 
free parameter $m$ with cosmological data sets which solely probe the 
background evolution of this model. Indeed, to the best of our knowledge, this 
is the first analysis which reports a constraint on the model parameter $m$. We 
were able to place some limits on $m$, since this parameter appears in the 
variation of the fine--structure constant $f(T)$ relationship defined by Eq. 
(\ref{f_T_fine_struc_const}).

We report the model parameter constraints  in the penultimate panel of Table 
\ref{tab:f_models}, and the marginalised two--dimensional likelihood 
constraints are depicted in Fig. \ref{fig:f4CDM_posteriors}. With the 
considered data sets, we were able to place a lower bound on 
$m\gtrsim0.57\,(m\gtrsim0.5)$ with the SN + CC + $H_0^R$ (SN + CC + KVNO/ SN + 
CC + KVNO + $H_0^R$) joint data set.

In order for this model to satisfy the adopted tight limits on the variation of 
the fine--structure constant, $\beta_F^{}$ was robustly constrained to 
$\sim10^{-6}$. Thus, the $\beta_F^{}$ constraints imposed by the KVNO data set 
were found to be of a similar order to the inferred constraints on the 
theoretical phenomenological parametrisations of section 
\ref{sec:phenom_cur_constraints}. Furthermore, this model seems to favor 
slightly low values of $\Omega^\mathrm{m}_0$, particularly when high $H_0^{}$ 
values are obtained. Consequently, this model will be disfavored in light of 
the $H_0^{}$ tension, as already highlighted in Refs.
\cite{Nesseris:2013jea,Xu:2018npu,Anagnostopoulos:2019miu}. 

\subsubsection{The $f_5(T)$ model}

Our last model is the hyperbolic--tangent model \cite{Wu:2010av} which is 
specified as follows
\begin{equation}
\label{fTmodel5}
    f_5(T)=-T+\alpha_5\, T^n\tanh\left(\frac{T_0}{T}\right)\,,
\end{equation}
with model parameters $\alpha_5$ and $n$. From Eq. (\ref{Friedmann_1}), we find 
that
\begin{equation}    
\alpha_5=\frac{\left(6H_0^2\right)^{1-n}\left(1-\Omega^\mathrm{m}_0\right)}{
(2n-1)\tanh(1)-2\sech^2(1)}\,,
\end{equation}
and therefore $n$ will be the varying model parameter. Similar to the previous 
$f_4(T)$ model,  the $\Lambda$CDM cosmology cannot be recovered as a limiting 
case
of the $f_5(T)$ model for any arbitrary value of $n$. Therefore, the parameter 
$n$ does not
characterise the deviation from the concordance model of cosmology.

\begin{figure}[t]
\begin{center}
    \includegraphics[width=0.88\columnwidth]{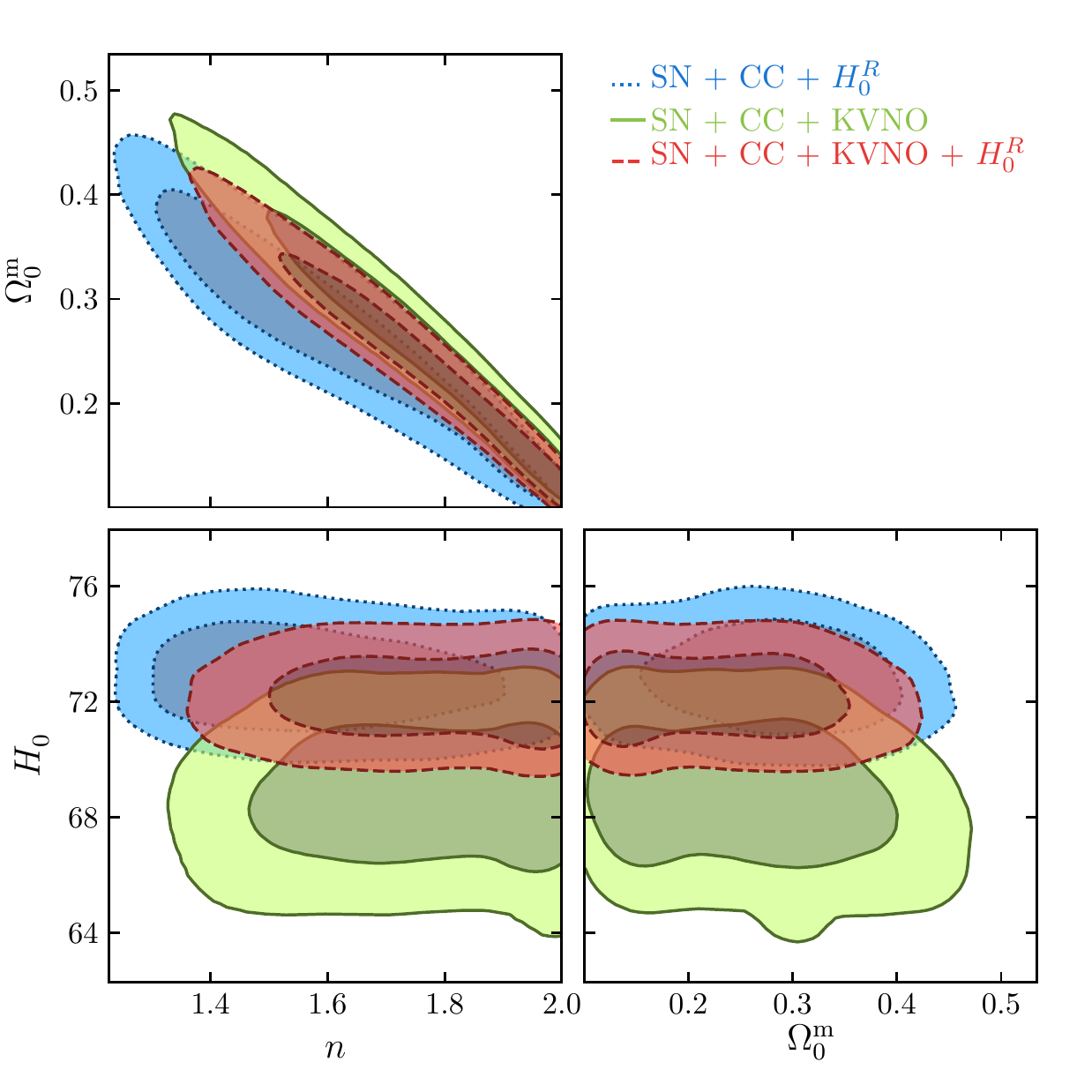}
    \caption{\label{fig:f5CDM_posteriors}{Marginalised two--dimensional likelihood constraints on the $f_5^{}(T)$ model parameters specified in Eq. (\ref{fTmodel5}).}}
\end{center}
\end{figure}

\begin{figure}[t]
\includegraphics[width=0.5\columnwidth]{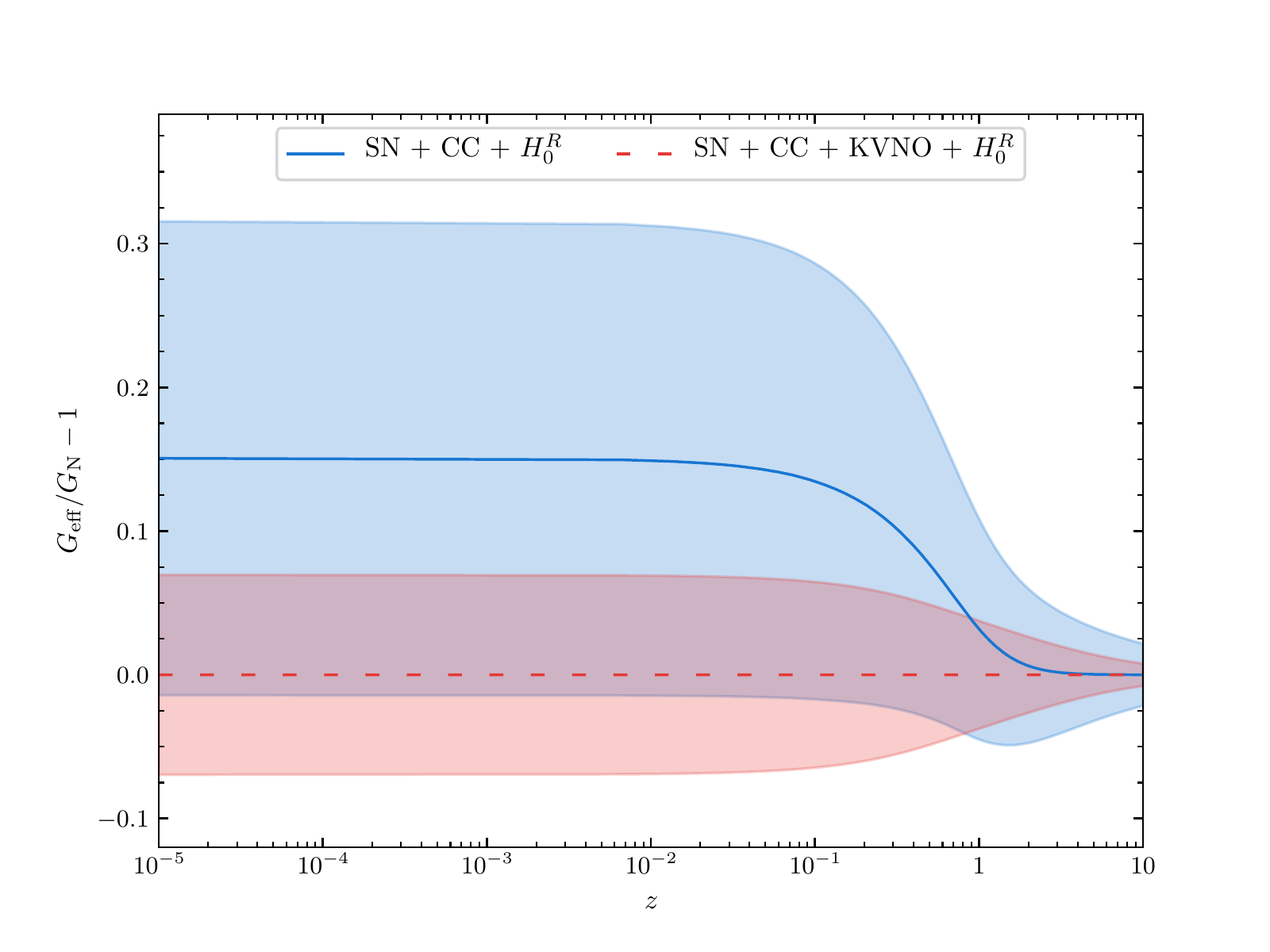}
\includegraphics[width=0.5\columnwidth]{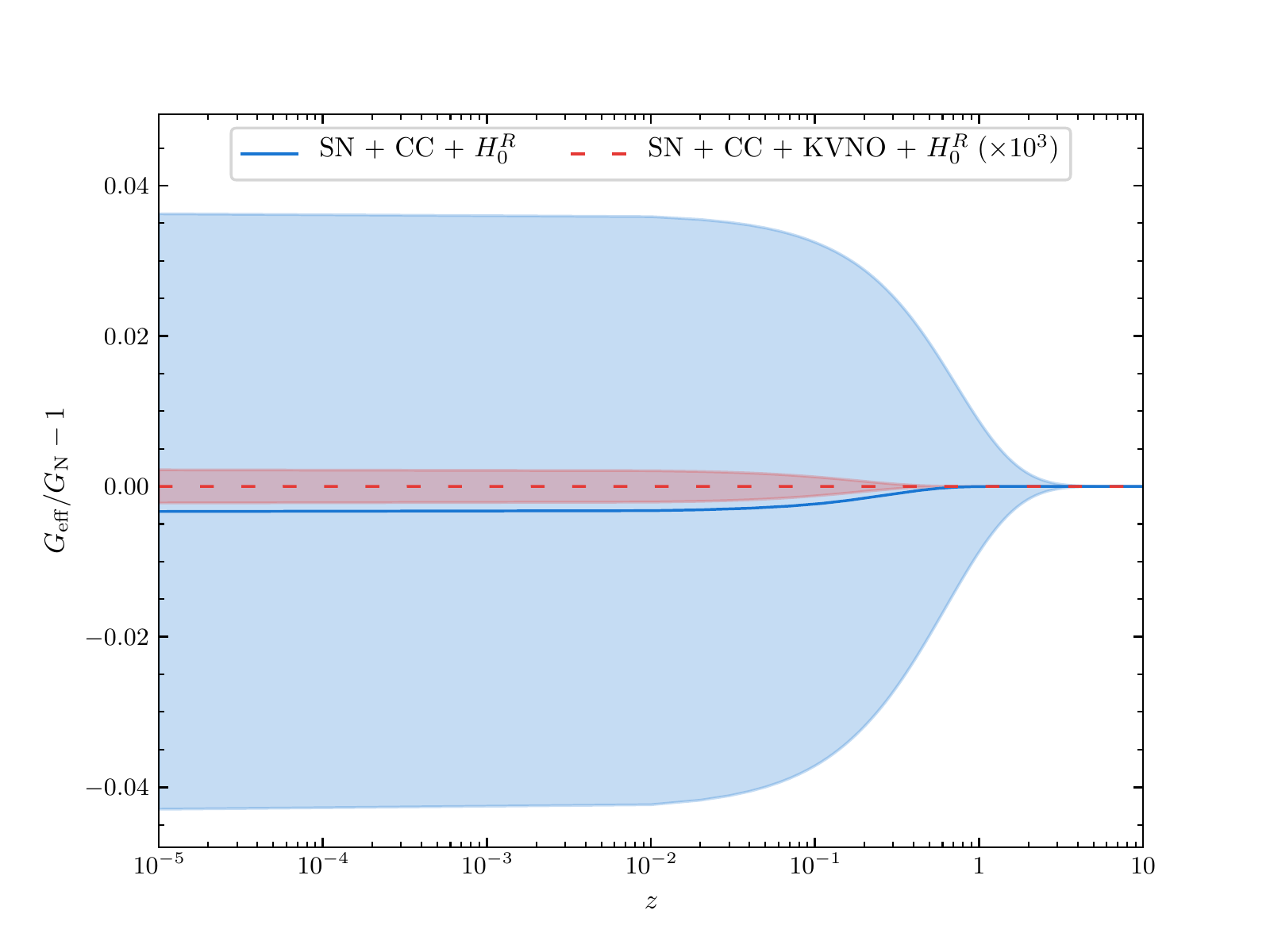}
\includegraphics[width=0.5\columnwidth]{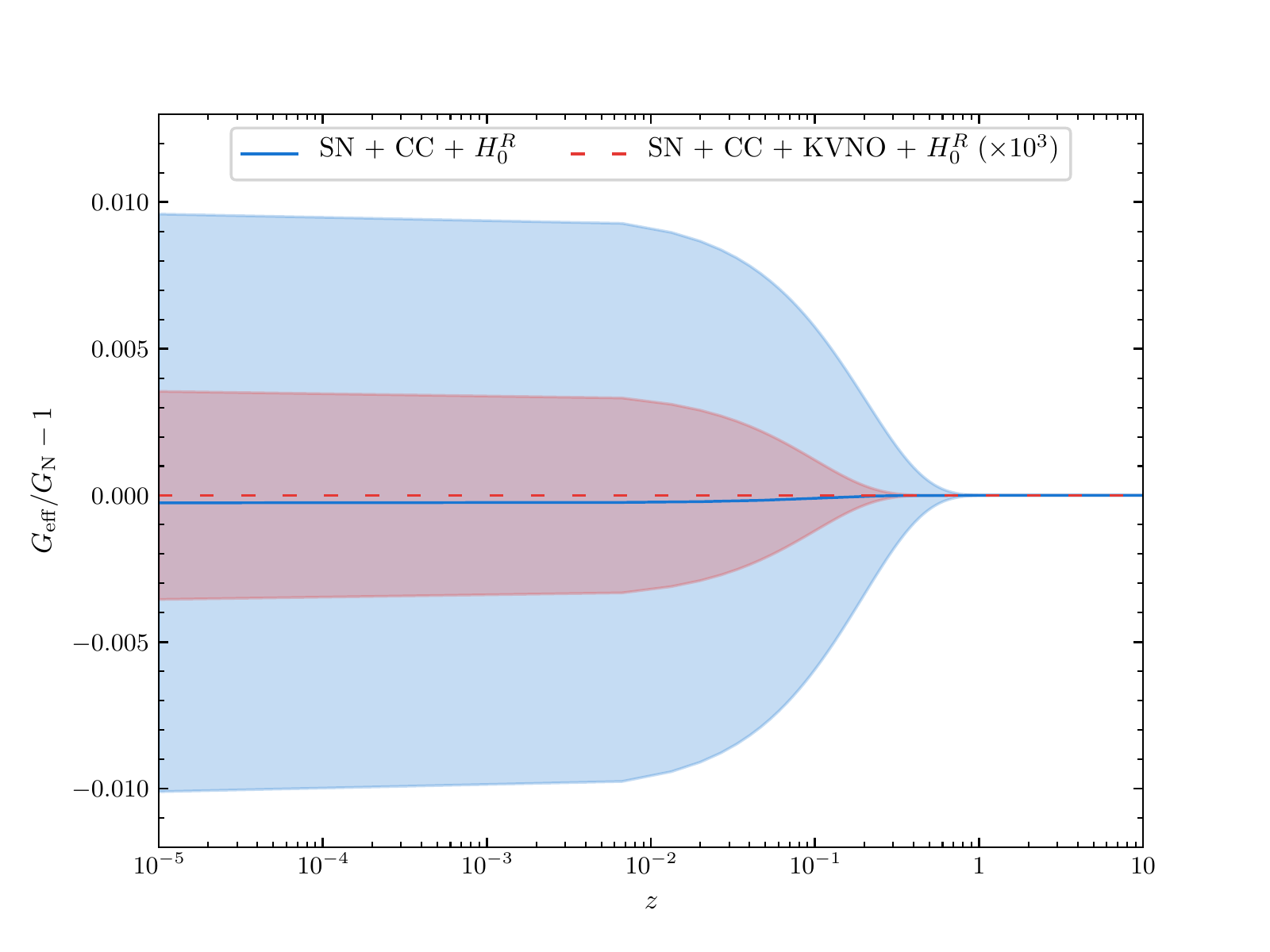}
\includegraphics[width=0.5\columnwidth]{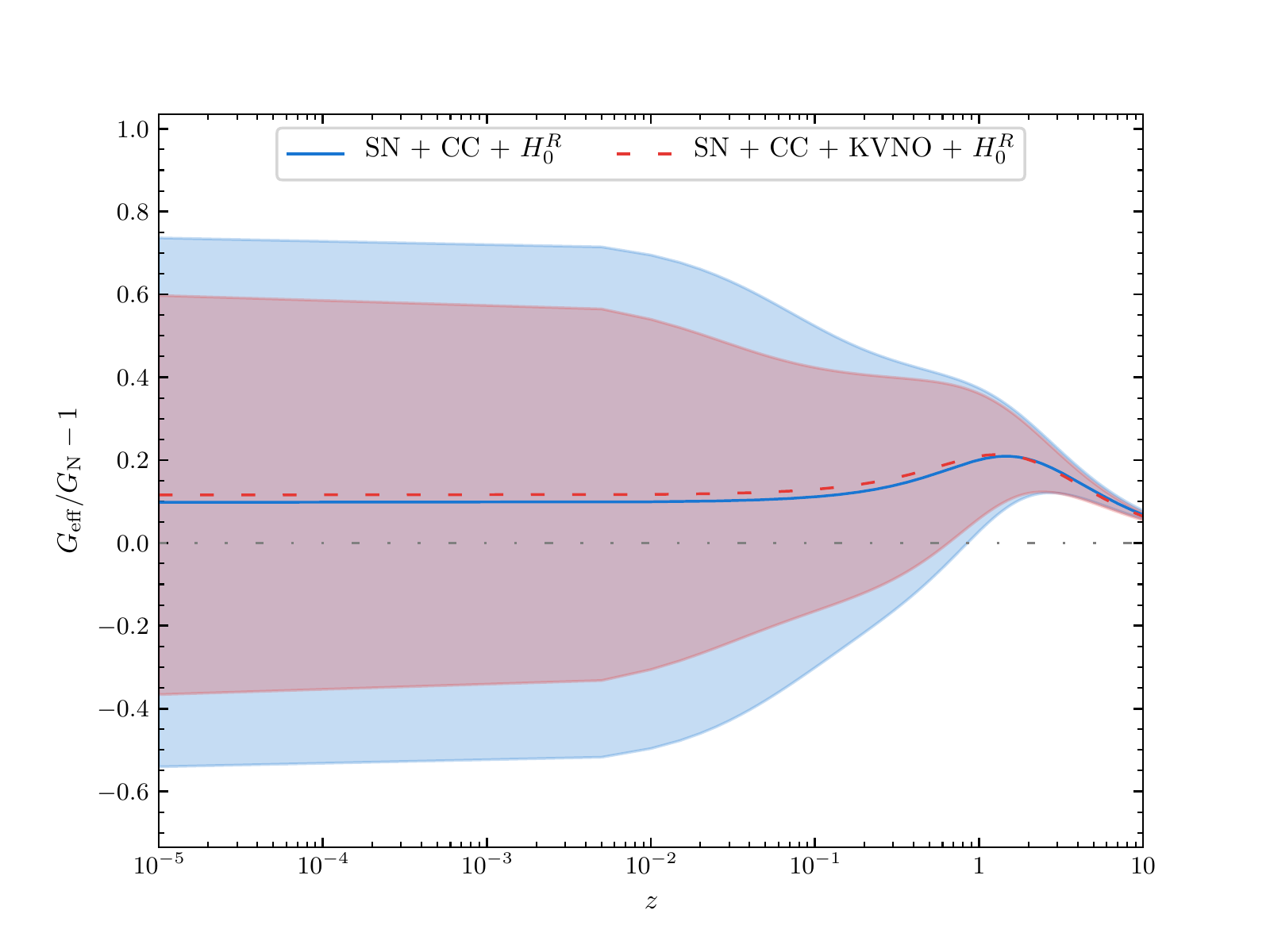}
\begin{center}
\includegraphics[width=0.5\columnwidth]{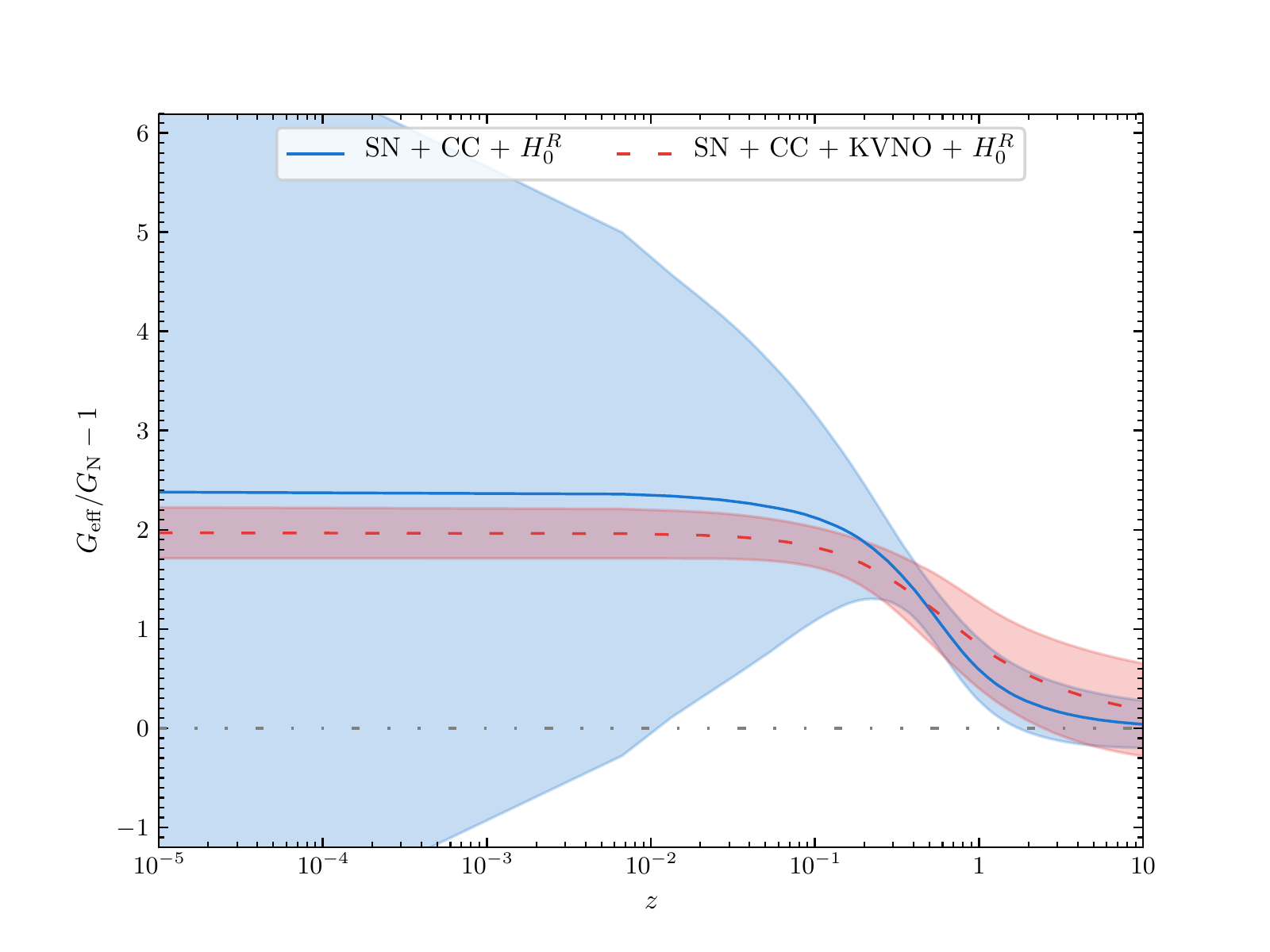}
\end{center}
\caption{\label{fig:G_eff}{Reconstruction of $G_\mathrm{eff}/G_\mathrm{N}-1$ and its $1\sigma$ uncertainty as a function of redshift, for the five $f(T)$ models considered in this work, namely $f_1^{}(T)$ (top left), $f_2^{}(T)$ (top right), $f_3^{}(T)$ (middle left), $f_4^{}(T)$ (middle right) and $f_5^{}(T)$ (bottom) models.}}
\end{figure}

The derived constraints are listed in the last panel of Table 
\ref{tab:f_models}, and the corresponding marginalised confidence contours are 
illustrated in Fig. \ref{fig:f5CDM_posteriors}. From the SN + CC + $H_0^R$ data 
set, we obtain a smaller value of $n$ when compared with the inferred mean 
value of this parameter from the other data sets which include the KVNO 
measurements. Given that small values of $n\lesssim1.69$ \cite{Wu:2010av} 
naturally give rise to the crossing of the phantom divide line, the KVNO data 
set restricts this possibility as higher values of $n$ are preferred. We note 
that our inferred constraints on $n$ agree with the reported results in Refs. 
\cite{Cardone:2012xq,Qi:2017xzl,Xu:2018npu,Anagnostopoulos:2019miu}, although 
in these analyses $\beta_F$ was neglected. We further observe that the higher 
the value of $n$, the smaller the value of $\Omega^\mathrm{m}_0$, which is 
consistent with Ref. \cite{Xu:2018npu,Anagnostopoulos:2019miu}. This will 
however make this model inconsistent with the CMB data.

Moreover, the $\beta_F^{}$ electromagnetic coupling parameter is loosely  
constrained with the SN + CC + $H_0^R$ joint data set, however the inclusion of 
the KVNO measurements lead to very stringent constraints on this parameter. 
Indeed, the inferred constraints on $\beta_F^{}$ with the SN + CC + KVNO and SN 
+ CC + KVNO + $H_0^R$ joint data sets are found to be similar to the derived 
constraints in the phenomenological parametrisations of section 
\ref{sec:phenom_cur_constraints} and in the case of the above $f_4(T)$ model. 
Consequently, the $f_5(T)$ hyperbolic--tangent model is also found to be in 
agreement with the distance--duality relation.

\subsection{\label{sec:gravitational_constant}Implications for an effective Newton's constant}

As already mentioned in section \ref{sec:fT_cosmology}, $f(T)$ gravity gives rise to an effective gravitational 
constant, in similarity with the majority of modified gravitational frameworks. This variation is generically given by 
\cite{Zheng:2010am,Nesseris:2013jea,Anagnostopoulos:2019miu} 
$G_\mathrm{eff}=\frac{G_N}{|f_T|}$, with $G_N$ being Newton's gravitational 
constant. Hence, the effective gravitational constant in the 
considered $f(T)$ models will coincide with $G_N$ at earlier times, and we 
expect some  deviation at   late--times.

In the panels of Fig. \ref{fig:G_eff}, we reconstruct the variation of  the 
effective gravitational constant, specified by the quantity 
$G_\mathrm{eff}/G_N-1$, at the 1$\sigma$ confidence level. For the power--law 
$f_1(T)$ model, the square--root--exponential $f_2(T)$ model and the exponential 
$f_3(T)$ model, we can clearly observe that $G_\mathrm{eff}\simeq G_N$, 
particularly when we make use of the SN + CC + KVNO + $H_0^R$ data set. Indeed, 
we observe that in these models, the KVNO measurements significantly restrict 
the deviation of $G_\mathrm{eff}$ from $G_N$, in agreement with current 
independent bounds on the time variation of the gravitational constant (see, for 
instance, Refs. 
\cite{GarciaBerro:2011wc,Williams:2004qba,Zhu:2018etc,Zhao:2018gwk,
Alvey:2019ctk} and references therein). Moreover, a much tighter 1$\sigma$ 
deviation of $G_\mathrm{eff}$ from $G_N$ is obtained at around $z\simeq1$, 
although this redshift is model dependent.

On the other hand, we observe a significant deviation of $G_\mathrm{eff}/G_N$ 
from unity at low redshifts in the logarithmic $f_4(T)$ and hyperbolic--tangent 
$f_5(T)$ scenarios. Such deviations are unequivocally forbidden by current 
constraints on the variation of Newton's gravitational constant 
\cite{GarciaBerro:2011wc,Williams:2004qba,Zhu:2018etc,Zhao:2018gwk,
Alvey:2019ctk}, and therefore we consider these models as cosmologically 
non-viable models, consistent with Refs. 
\cite{Nesseris:2013jea,Xu:2018npu,Anagnostopoulos:2019miu}. Indeed, only the 
$f(T)$ models which posses the $\Lambda$CDM model as a limiting case are in 
agreement with the condition of $G_\mathrm{eff}/G_N\simeq1$, which could 
therefore be considered as viable cosmological models.

\section{\label{sec:conc}Conclusions}

In this work, we focused our attention on the $f(T)$ gravitational framework in section 
\ref{sec:f_T}. After our concise discussion on TG and $f(T)$ cosmology, we   
focused on the induced variation of the fine--structure constant in the $f(T)$ 
gravitational scenario. Indeed, we have revisited and updated the theoretical 
$f(T)$ relationship of $\Delta\alpha/\alpha$ in Eq. 
(\ref{f_T_fine_struc_const}), and the modification of the luminosity distance in 
Eq. (\ref{eq:dL_f_T}). 

In section \ref{sec:f_T_alpha}  we  proceeded to the confrontation of five 
$f(T)$ models with the Supernovae Type Ia Pantheon Sample, Hubble parameter 
measurements and measurements of the variation of the fine--structure constant. 
We have considered three models $(f_1(T),\,f_2(T),\,f_3(T))$ which posses the 
$\Lambda$CDM model as a limiting case. From our inferred results of Table 
\ref{tab:f_models}, we observe that the $f_1(T),\,f_2(T),$ and $f_3(T)$ models 
do not exclude the $\Lambda$CDM paradigm. It was also found that these are 
cosmologically viable models, since the reconstructed deviation of  their 
effective gravitational constant from $G_N$ is negligible and in an excellent 
agreement with current experimental bounds.

On the other hand, the remaining $f(T)$ models do not contain the $\Lambda$CDM 
scenario as a particular limit. The logarithmic $f_4(T)$ model is identical to 
the spatially flat self--accelerating branch of the DGP model at the background 
level, and therefore we expected that this model will not be cosmologically 
viable. Indeed, the reconstructed evolution of $G_\mathrm{eff}/G_N$ 
significantly deviated from unity at late--times, which clearly is not 
consistent with current bounds on $G_\mathrm{eff}/G_N$. The remaining 
hyperbolic--tangent model is characterised by the crossing of the phantom divide 
line, although our inferred constraints did not favor this possibility due to a 
preference to relatively large values of the $f_5(T)$ model parameter $n$. 
Moreover, the reconstructed evolution of $G_\mathrm{eff}/G_N$ was not found to 
be in agreement with the respective experimental bounds.

Also in appendix \ref{sec:phenom}, we explore a number of widely known theoretical parametrisations of the violation of the distance--duality relation, on which we imposed very stringent constraints 
$(\mathcal{O}(10^{-7}))$ by adopting several measurements of the variation of the fine--structure constant. From this analysis, we clearly illustrated that current data sets are in an excellent agreement with the distance--duality relation irrespective of the adopted phenomenological parametrisation.

A common feature of all $f(T)$ models is that they are all in an excellent 
agreement with the distance--duality relation. Thus, with current measurements 
of the variation of the fine--structure constant, we have been able to 
confirm  the validity of EEP in $f(T)$ gravity. We   expect that 
the relevant constraints will significantly improve in the era of the new 
generation
of high--resolution ultra--stable spectrographs, such as
ESPRESSO \cite{1750848} and ELT--HIRES \cite{10.1117/12.2231653}, which will 
lead to improvements in local atomic clock tests and complimentary cosmological 
observations.

\begin{acknowledgments}
The authors would like to acknowledge funding support Cosmology@MALTA  which is 
supported by the University of Malta. The authors would like to acknowledge 
networking support by the COST Action CA18108. The authors would also like to 
acknowledge networking support by the COST Action GWverse CA16104. JM would like to acknowledge``IPAS - 2018-011: TeleGravity" project for travel funds which was funded by \textit{The Malta Council for Science and Technology (MCST)}.
\end{acknowledgments}

\appendix
\section{\label{sec:phenom}Phenomenological violation of the cosmic distance--duality relation}
The EEP could be easily broken by introducing 
a phenomenological nonminimal multiplicative coupling between a scalar field 
$\phi$  and matter fields. For instance, in the electromagnetic sector, the 
action formalism would be given by
\begin{equation}
    \mathcal{S}_\mathrm{EM}^{} = \int \mathrm{d}^4x\sqrt{-g}\,B_F^{}(\phi)\,\mathcal{L}_\mathrm{EM}^{}\;,
\end{equation}
where the electromagnetic Lagrangian is denoted by $\mathcal{L}_\mathrm{EM}^{}$, 
$g$ is the determinant of the space--time metric $g_{\mu\nu}$, and the 
scalar field dependent electromagnetic coupling is denoted by $B_F^{}(\phi)$. 
We remark that the dynamical evolution of the scalar field and the metric 
tensor are not relevant at this point, and such dynamics are encoded in the 
scalar--gravitational field Lagrangian.

After the variation of the above action 
with respect to the electromagnetic four--potential $A^\mu$, we arrive at the 
homogeneous modified Maxwell equations
\begin{equation}
    \label{eq:Maxwell}
    \mathring{\nabla}_\nu\left(B_F^{}(\phi)\,F^{\mu\nu}\right) = 0\;,
\end{equation}
with $F_{\mu\nu}=\partial_\mu A_\nu-\partial_\nu A_\mu$ being  the standard 
antisymmetric Faraday tensor, and $\mathring{\nabla}_\nu$ being the regular 
covariant derivative calculated with the Levi-Civita connection. From Eq. 
(\ref{eq:Maxwell}) we know that photons propagate on null geodesics, and 
therefore the reciprocity relation still holds \cite{Ellis:1971pg}. However, the 
number of photons is no longer conserved, which consequently leads to a 
violation of Etherington's relation \cite{1933PMag...15..761E}. Thence, one 
could parametrise the violation of the distance--duality relation by
\begin{equation}
    \eta(z) = \frac{D_L^{}(z)}{D_A^{}(z)\,(1 + z)^2}\;,
\end{equation}
where $D_L^{}(z)$ and $D_A^{}(z)$ are the  luminosity distance and angular 
diameter distance at redshift $z$, respectively. Clearly, the distance--duality 
relation is recovered when $\eta(z)=1$.

 In our analyses, we will be adopting the following commonly used 
phenomenological  parametrisations
 \begin{align}
     \eta(z) &= \eta_0^{}\;,\vphantom{\frac11}\label{eq:eta_0}\\
     \eta(z) &= 1 + \eta_1^{}z\;,\vphantom{\frac11}\label{eq:eta_1}\\
     \eta(z) &= 1 + \eta_2^{}\frac{z}{1 + z}\;,\vphantom{\frac11}\label{eq:eta_2}\\
     \eta(z) &= 1 + \eta_3^{}\ln(1 + z)\;,\vphantom{\frac11}\\
     \eta(z) &= (1 + z)^\epsilon\;.\vphantom{\frac11}\label{eq:epsilon}
 \end{align}
The parametrisation of Eq. (\ref{eq:eta_0}) is simply considering a 
time--independent constant parameter which is not necessarily equal to unity 
\cite{Uzan:2004my,DeBernardis:2006ii}, whereas the second theoretical 
parametrisation \cite{Holanda:2010ay} is depicting a Taylor expansion at low 
redshifts which is ill behaved at high redshifts. The divergence problem in Eq. 
(\ref{eq:eta_1}) is fixed in the parametrisation of Eq. (\ref{eq:eta_2}) 
\cite{Holanda:2010ay}, while the fourth parametrisation appears in dilaton--type 
models \cite{Martins:2017yxk,Damour:2002mi,Damour:2002nv}. The last theoretical 
parametrisation of Eq. (\ref{eq:epsilon}) was studied in the context of 
deviations from cosmic transparency \cite{Avgoustidis:2009ai,Nair:2012dc}, which 
could arise from astrophysical attenuation processes as well as from exotic 
physics including photon--axion mixing 
\cite{Bassett:2003vu,Avgoustidis:2010ju,Tiwari:2016cps}. 

Several studies have constrained  the theoretical parametrisations presented in 
Eqs. (\ref{eq:eta_0}) -- (\ref{eq:epsilon}), in which different kinds of 
cosmological and local probes have been adopted (see, for instance, Refs. 
\cite{Uzan:2004my,DeBernardis:2006ii,Lazkoz:2007cc,
Holanda:2010ay,Holanda:2011hh,Holanda:2010vb,Li:2011exa,
Liang:2011gm,Meng:2011nt,Cao:2011fw,Cardone:2012vd, 
Holanda:2012at,Yang:2013coa,Li:2013cva,Holanda:2014lna,Hees:2014lfa, 
Costa:2015lja,Lv:2016mmq,Ma:2016bjt,Lin:2018qal,Ruan:2018dls,Goncalves:2019xtc} 
and references therein), and no significant deviation from the distance--duality 
relation has not been reported yet. Consequently, the likelihoods of 
$\eta_0^{}-1,\,\eta_1^{},\,\eta_{2}^{},\,\eta_3^{},$ and $\epsilon$ are expected 
to peak at zero in order to satisfy the distance--duality relation.  

\begin{table*}
    \centering
    \setlength\extrarowheight{7pt}
    \addtolength{\tabcolsep}{-1.48pt}
    \begin{tabular}{ c c c c c c c}
        \hline
        \hline
        {Data set} & {$(\eta_0^{}-1)\,\left[\times10^{-7}\right]$} & 
{$\eta_1^{}\,\left[\times10^{-7}\right]$} & 
{$\eta_2^{}\,\left[\times10^{-7}\right]$} & 
{$\eta_3^{}\,\left[\times10^{-7}\right]$} & 
{$\epsilon\,\left[\times10^{-7}\right]$} \\[.5em]
		\hline
		{K} & {$-28.6^{+5.8}_{-5.7}$} & {$-18.6^{+3.9}_{-3.9}$} & 
{~$-52.2^{+10.6}_{-9.9}$} & {$-33.5^{+6.8}_{-6.6}$} & {$-33.4^{+6.8}_{-6.7}$}  
\\
		{V} & {~~$10.5^{+6.2}_{-6.3}$} &  {~~~~$8.4^{+3.5}_{-3.6}$} & 
{~~~$20.0^{+10.0}_{-10.0}$} & {~~$13.9^{+6.3}_{-6.4}$} & 
{~~$13.8^{+6.3}_{-6.4}$} \\ 
		{N} & {~$-3.0^{+3.5}_{-3.3}$} & {~$-1.5^{+2.4}_{-2.5}$} & 
{~$-4.5^{+5.8}_{-6.1}$} & {~$-2.8^{+3.9}_{-4.0}$} & {~$-2.8^{+3.9}_{-4.0}$} \\ 
		{KV} & {$-10.8^{+4.2}_{-4.3}$} & {~$-3.7^{+2.6}_{-2.6}$} & 
{$-15.6^{+7.3}_{-7.3}$} & {~$-8.5^{+4.6}_{-4.8}$} & {~$-8.4^{+4.5}_{-4.7}$} \\ 
		{NO} & {~~~$0.00^{+0.30}_{-0.31}$} & {~$-0.6^{+1.6}_{-1.7}$} & 
{~$-0.5^{+2.3}_{-2.3}$} & {~$-0.5^{+2.0}_{-2.0}$} & {~$-0.6^{+2.0}_{-2.0}$} \\ 
		{KVN} & {~$-6.0^{+2.6}_{-2.6}$} & {~$-2.5^{+1.8}_{-1.8}$} & 
{~$-9.2^{+4.7}_{-4.5}$} & {~$-5.2^{+2.9}_{-3.1}$} & {~$-5.3^{+3.0}_{-3.0}$} \\ 
		{NO + lab + $H_0^R$} & {~~~$0.00^{+0.31}_{-0.30}$} & 
{~~~~$0.3^{+1.1}_{-1.1}$} & {~~~$0.6^{+1.3}_{-1.3}$} & 
{~~~~$0.5^{+1.2}_{-1.2}$~} & {~~~$0.5^{+1.2}_{-1.2}$} \\ 
		{KVNO + lab + $H_0^R$} & {~$-0.06^{+0.31}_{-0.30}$} & 
{~$-0.3^{+1.0}_{-1.0}$} & {~~~$0.1^{+1.3}_{-1.2}$} & {~~$-0.1^{+1.2}_{-1.2}$~} & 
{~$-0.1^{+1.2}_{-1.2}$} \\[.5em] 
		\hline
		\hline
    \end{tabular}
    \caption{We report the mean value and the 68\% limits for each $\eta(z)$ parametrisation, as described in Eqs. (\ref{eq:eta_0}) -- (\ref{eq:epsilon}). The data sets are discussed in section \ref{sec:alpha_data}.}
    \label{tab:eta_models}
\end{table*}

\subsection{\label{sec:phenom_exp_constraints}Induced variation of  the 
fine--structure constant}

A number of well--known cosmological consequences arising  from the violation of 
the distance--duality relationship have been widely explored in the literature. 
We will be particularly interested in the induced variation of the 
electromagnetic fine--structure constant, which is explicitly related with the 
nonminimal electromagnetic coupling via $\alpha\propto B_F^{-1}(\phi)$ 
\cite{Damour:1994zq,Damour:1994ya,Damour:2002mi,Damour:2002nv}. Indeed, the 
unequivocal relationship of the redshift evolution of $\alpha(z)$, with the 
nonminimal electromagnetic coupling and Etherington's parameter $\eta(z)$, can 
be expressed as follows
\begin{equation}
    \label{eq:alpha_relation}
    \frac{\Delta\alpha(z)}{\alpha} \equiv \frac{\alpha(z) -  
\alpha_0^{}}{\alpha_0^{}} = \frac{B_F^{}\left(\phi_0^{}\right)}{B_F^{}(\phi)} - 
1 = \eta^2(z) - 1\;,
\end{equation}
where a 0--subscript indicates the current epoch values at $z=0$.  Thus, 
constraints on $\Delta\alpha(z)/\alpha$ can be interchanged to constraints on 
$\eta(z)$, and vice versa. Furthermore, the current temporal variation of 
$\alpha(z)$, simplifies to the following equation
\begin{equation}
    \frac{\dot{\alpha}}{\alpha}\bigg\vert_0  = 
-2H_0^{}\frac{\mathrm{d}\eta}{\mathrm{d}z}\bigg\vert_0\;,
\end{equation}
where the Hubble constant is denoted by 
$H_0^{}=a^{-1}\mathrm{d}a/\mathrm{d}t\vert_0=a^{-1}\dot{a}\vert_0$, with $t$ 
being the cosmic time and $a(t)$ is the cosmic scale factor of a spatially--flat 
Friedmann--Lema\^{i}tre--Robertson--Walker (FLRW) metric with $a_0^{}=1$.

\begin{figure*}
\includegraphics[width=0.5\columnwidth]{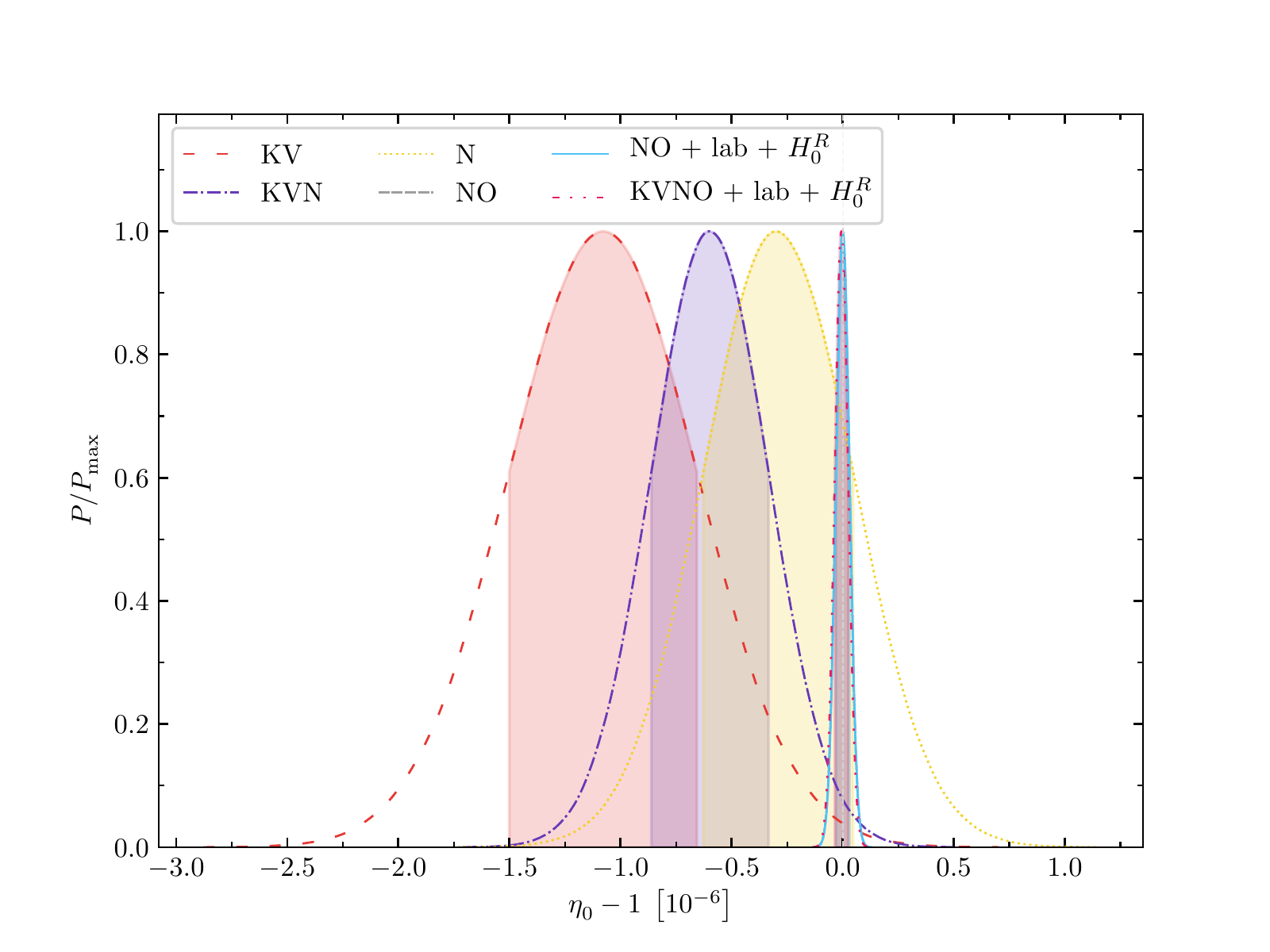}
\includegraphics[width=0.5\columnwidth]{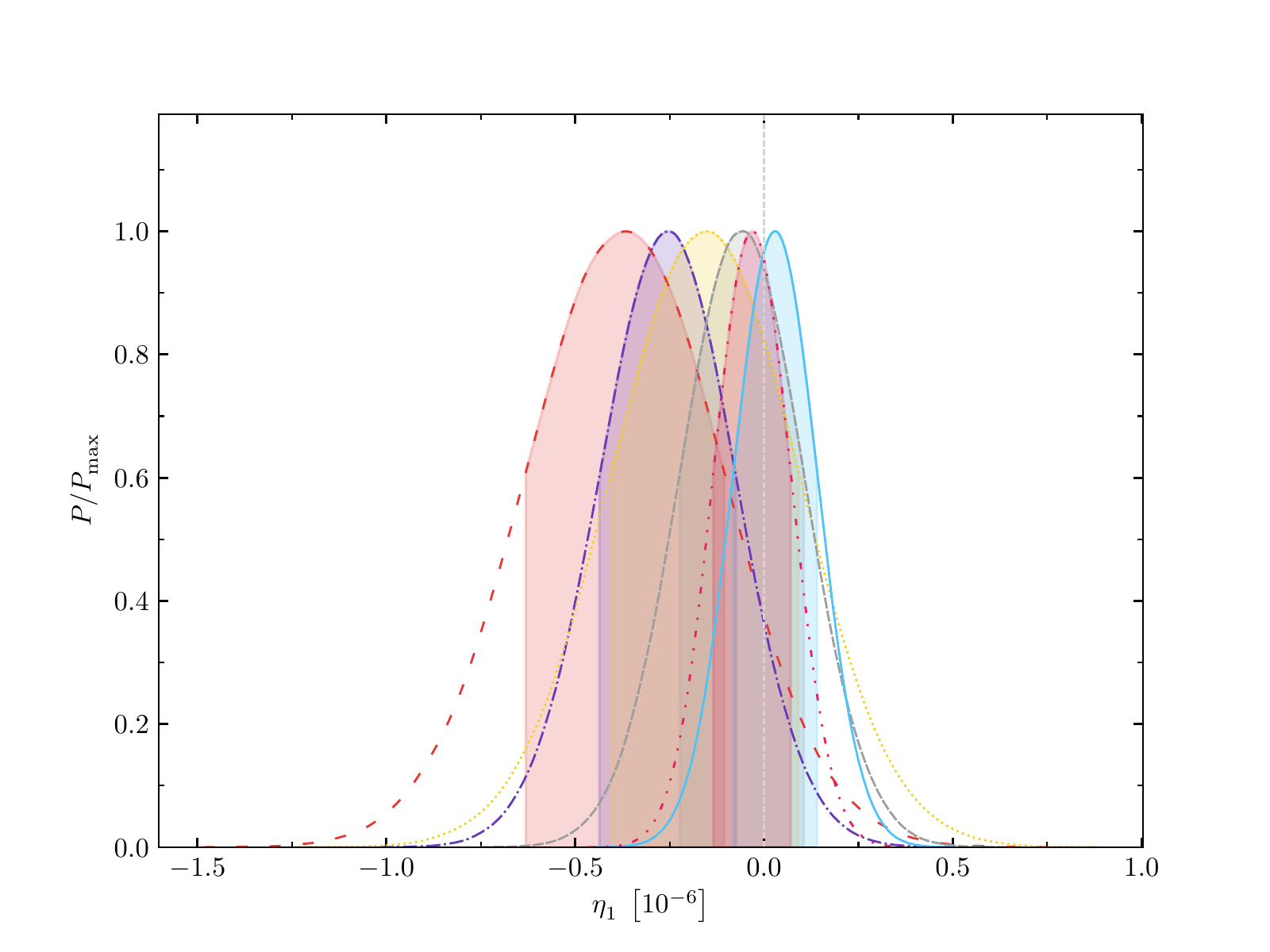}
\includegraphics[width=0.5\columnwidth]{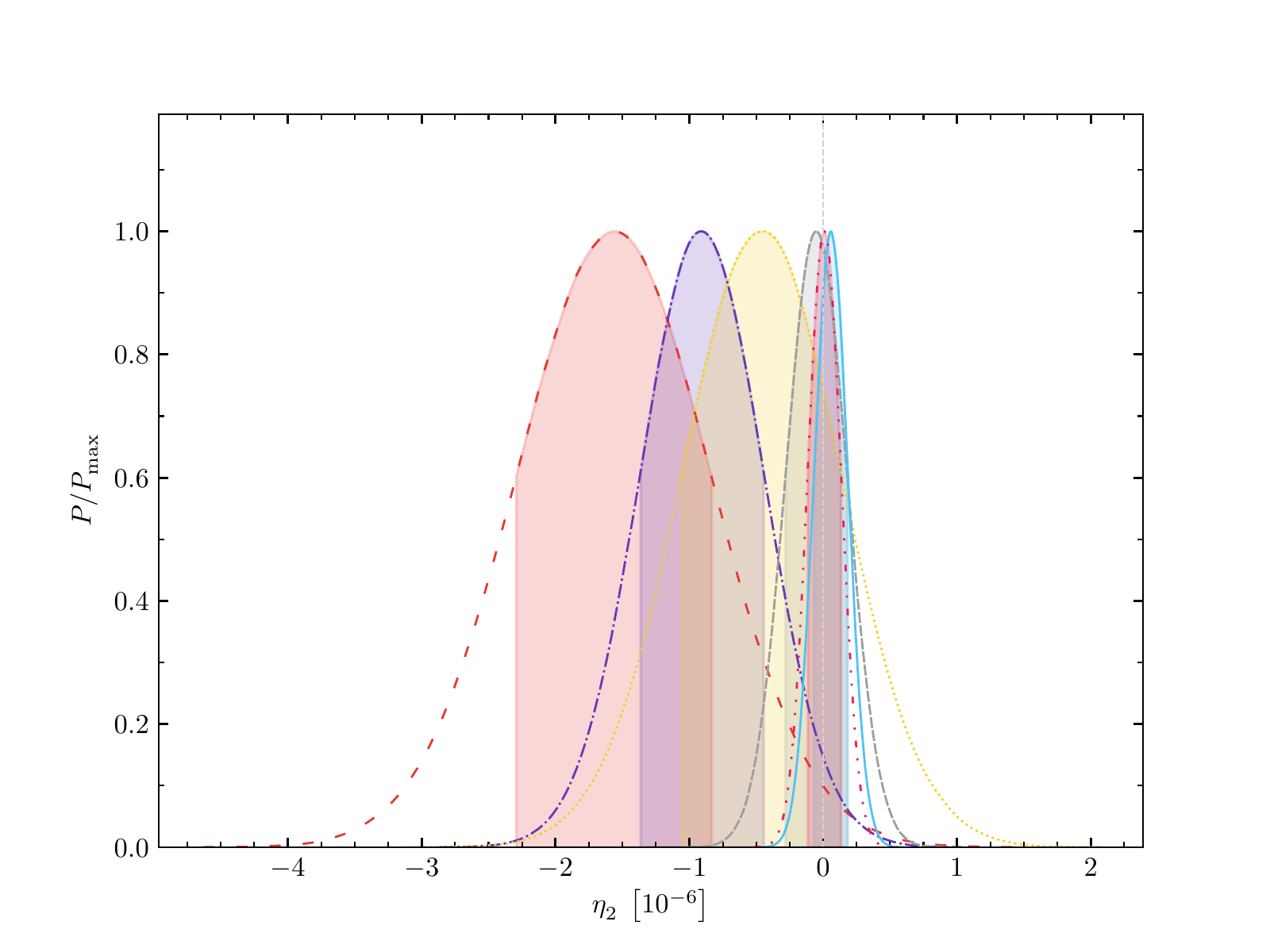}
\includegraphics[width=0.5\columnwidth]{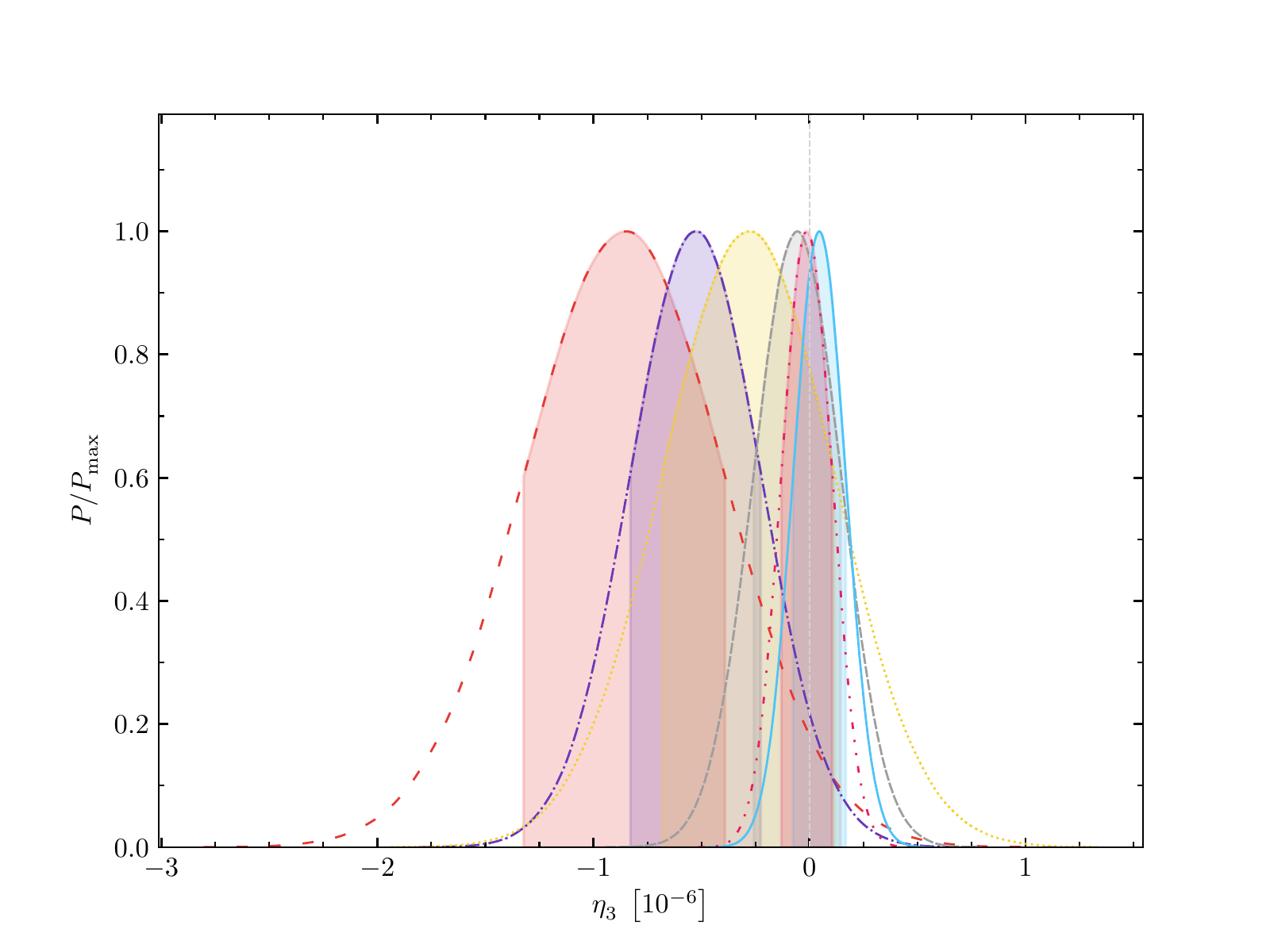}
\begin{center}
\includegraphics[width=0.5\columnwidth]{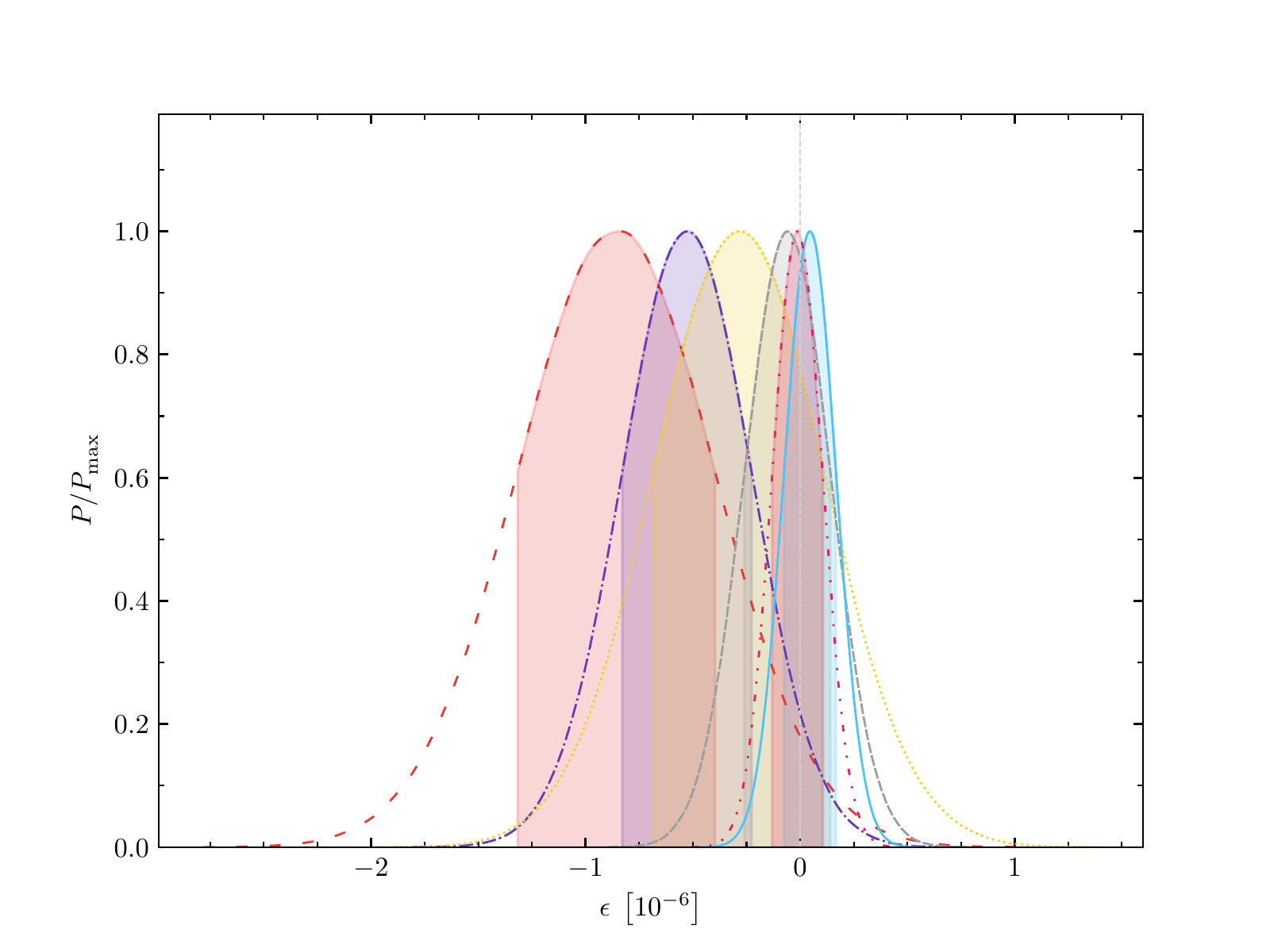}
\end{center}
\caption{\label{fig:eta_mar_posteriors}{Marginalised posterior distributions of the parameters characterising the phenomenological $\eta(z)$ theoretical parametrisations of Eqs. (\ref{eq:eta_0}) -- (\ref{eq:epsilon}).}}
\end{figure*}

\subsection{\label{sec:alpha_data}Data sets and methodology}

We will be implementing the   
methodology   outlined in section \ref{sec:f_T_alpha} and apply it for each phenomenological 
parametrisation of Eqs. (\ref{eq:eta_0}) -- (\ref{eq:epsilon}). We thus make use of the MCMC Ensemble sampler \texttt{emcee} 
\cite{ForemanMackey:2012ig}, and analyse our chains by the publicly available 
package \texttt{ChainConsumer} \cite{Hinton2016}.

\begin{figure*}[t]
\includegraphics[width=0.5\columnwidth]{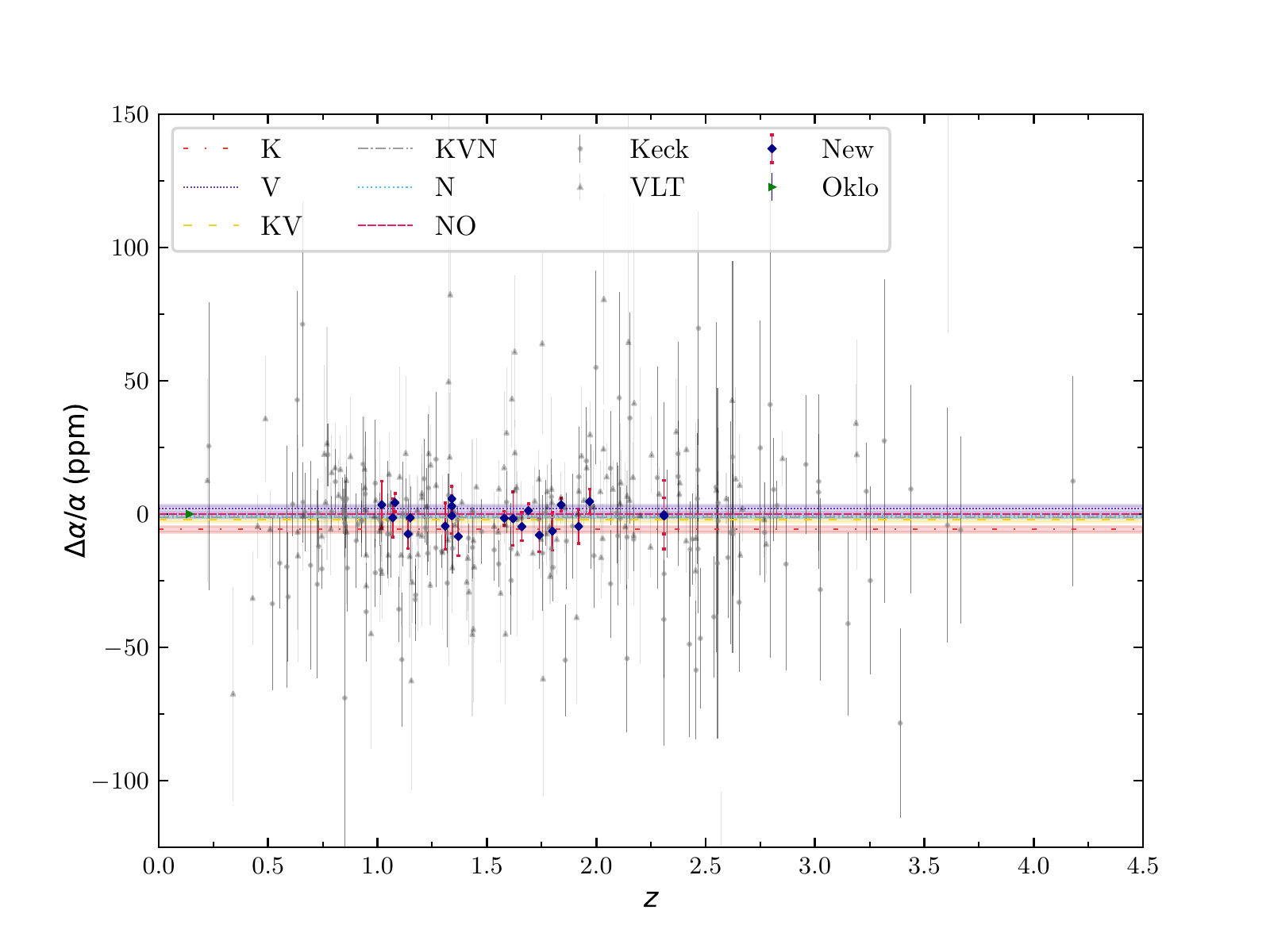}
\includegraphics[width=0.5\columnwidth]{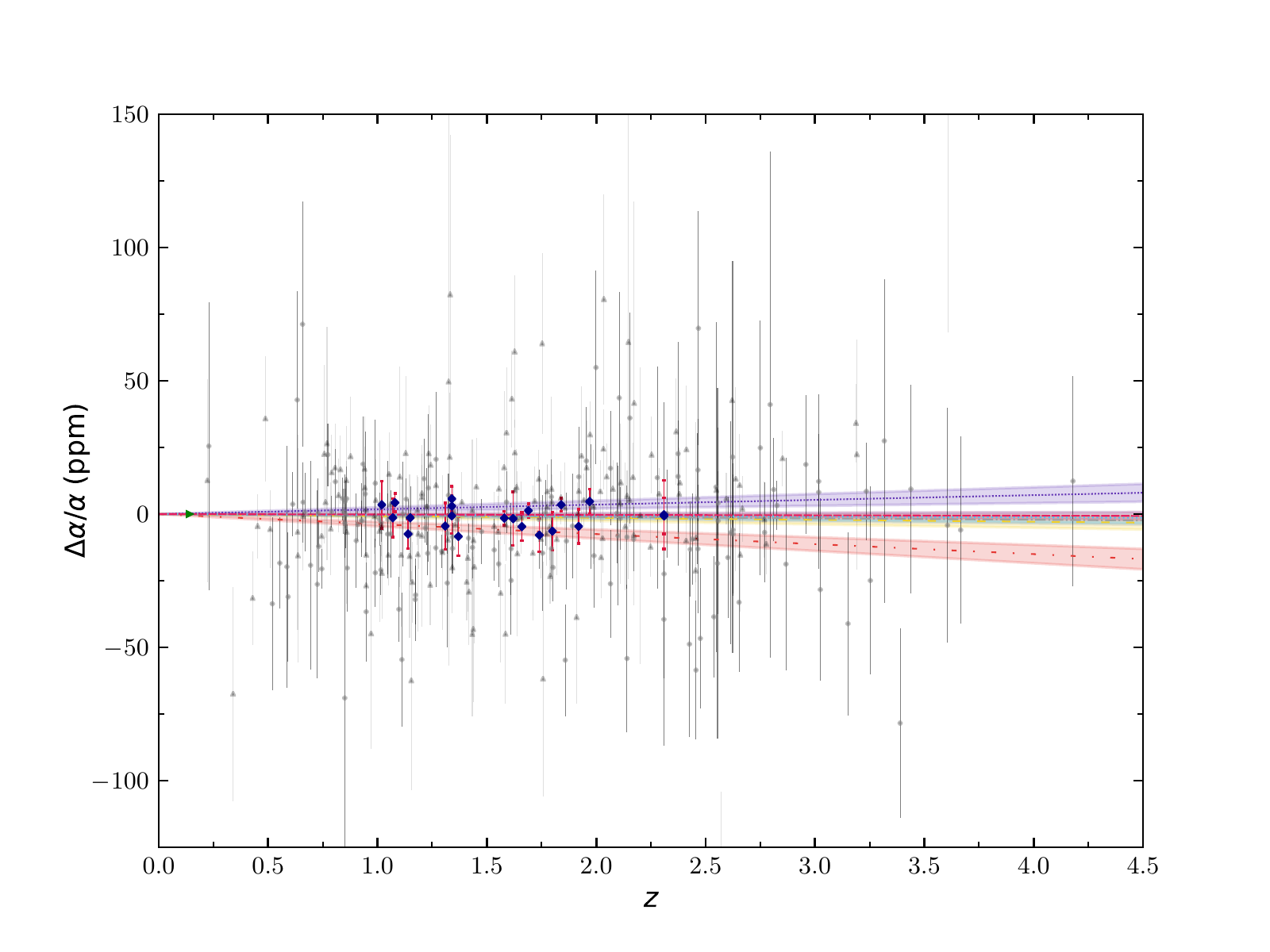}
\includegraphics[width=0.5\columnwidth]{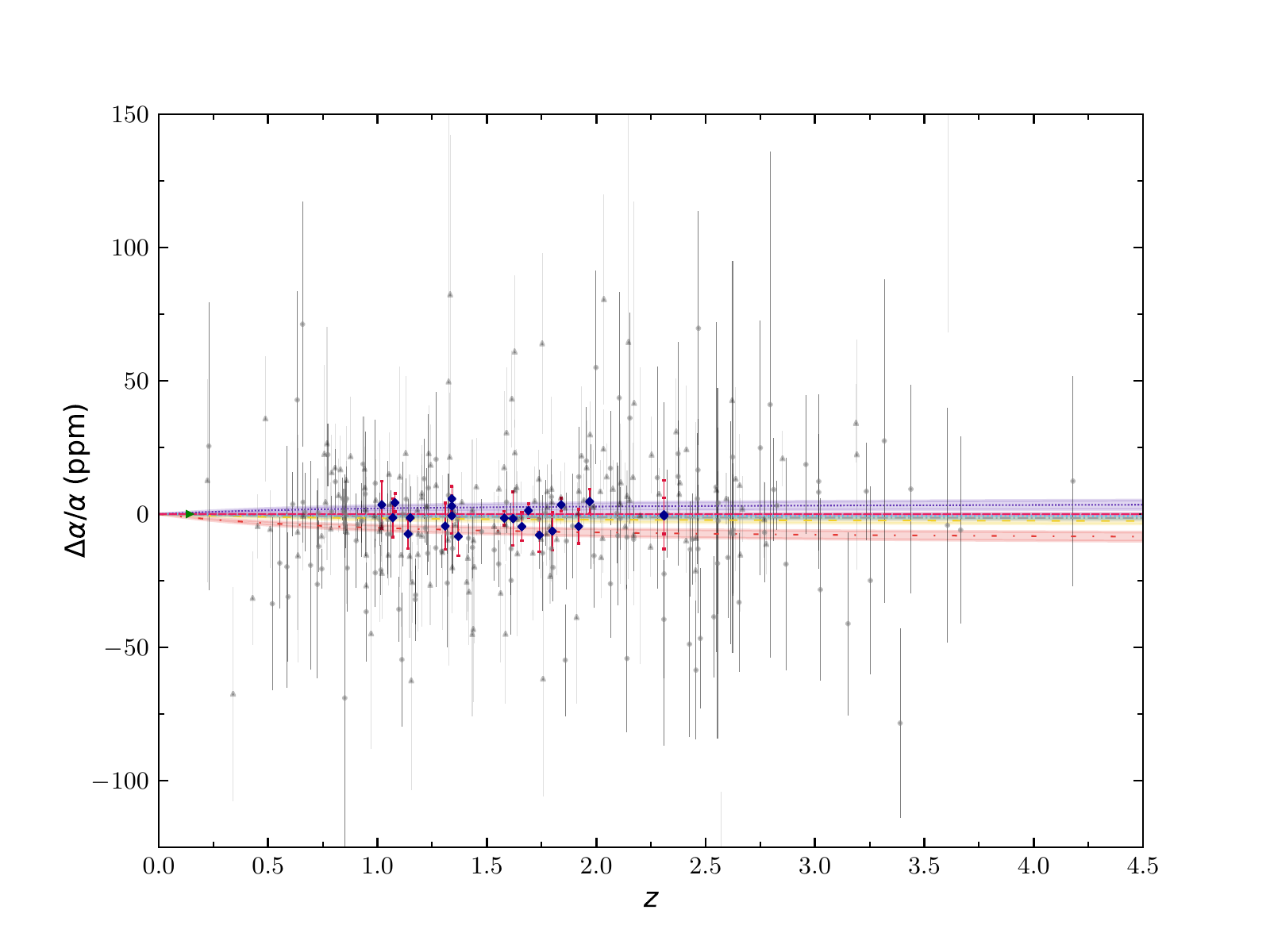}
\includegraphics[width=0.5\columnwidth]{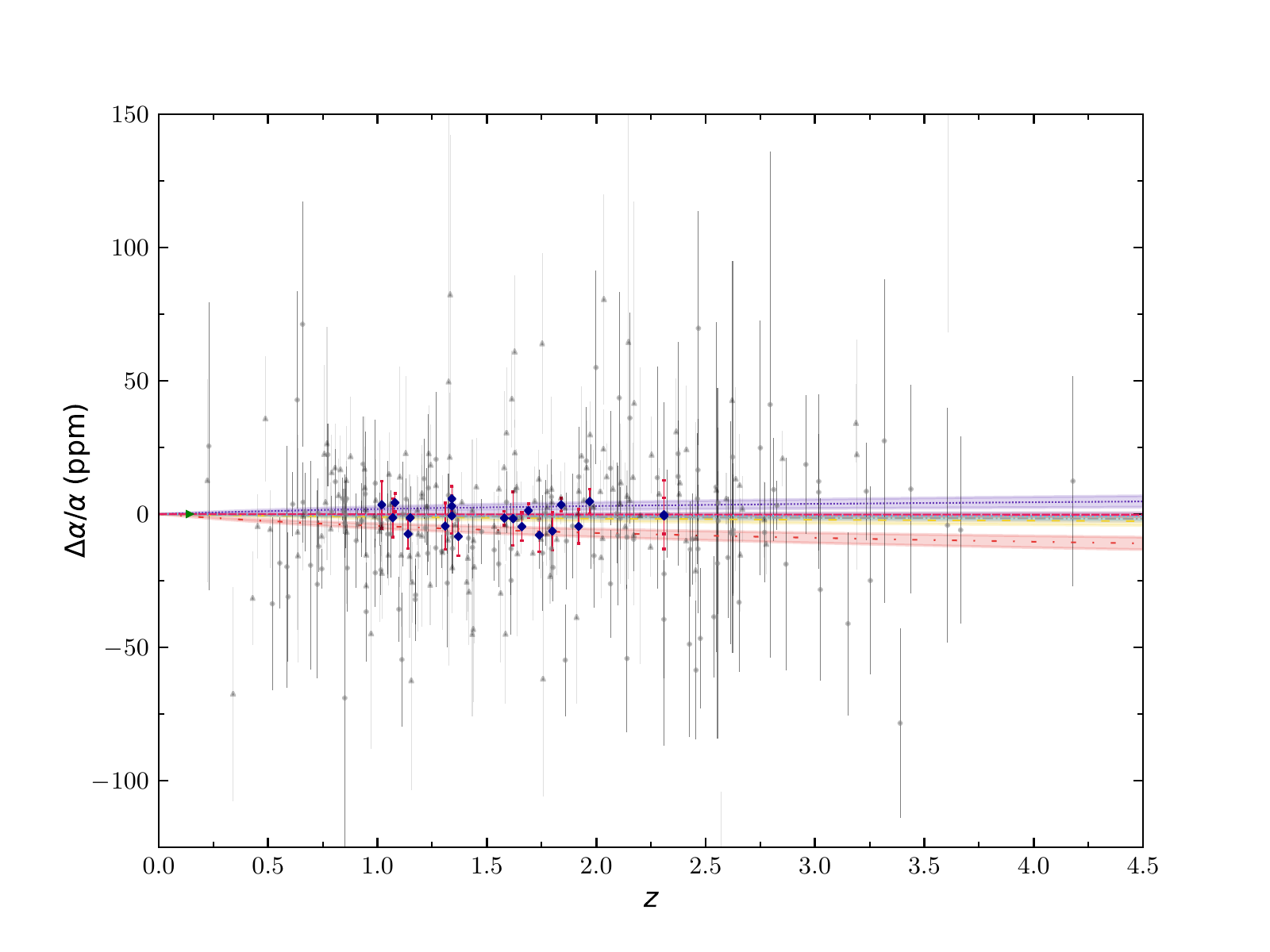}
\begin{center}
\includegraphics[width=0.5\columnwidth]{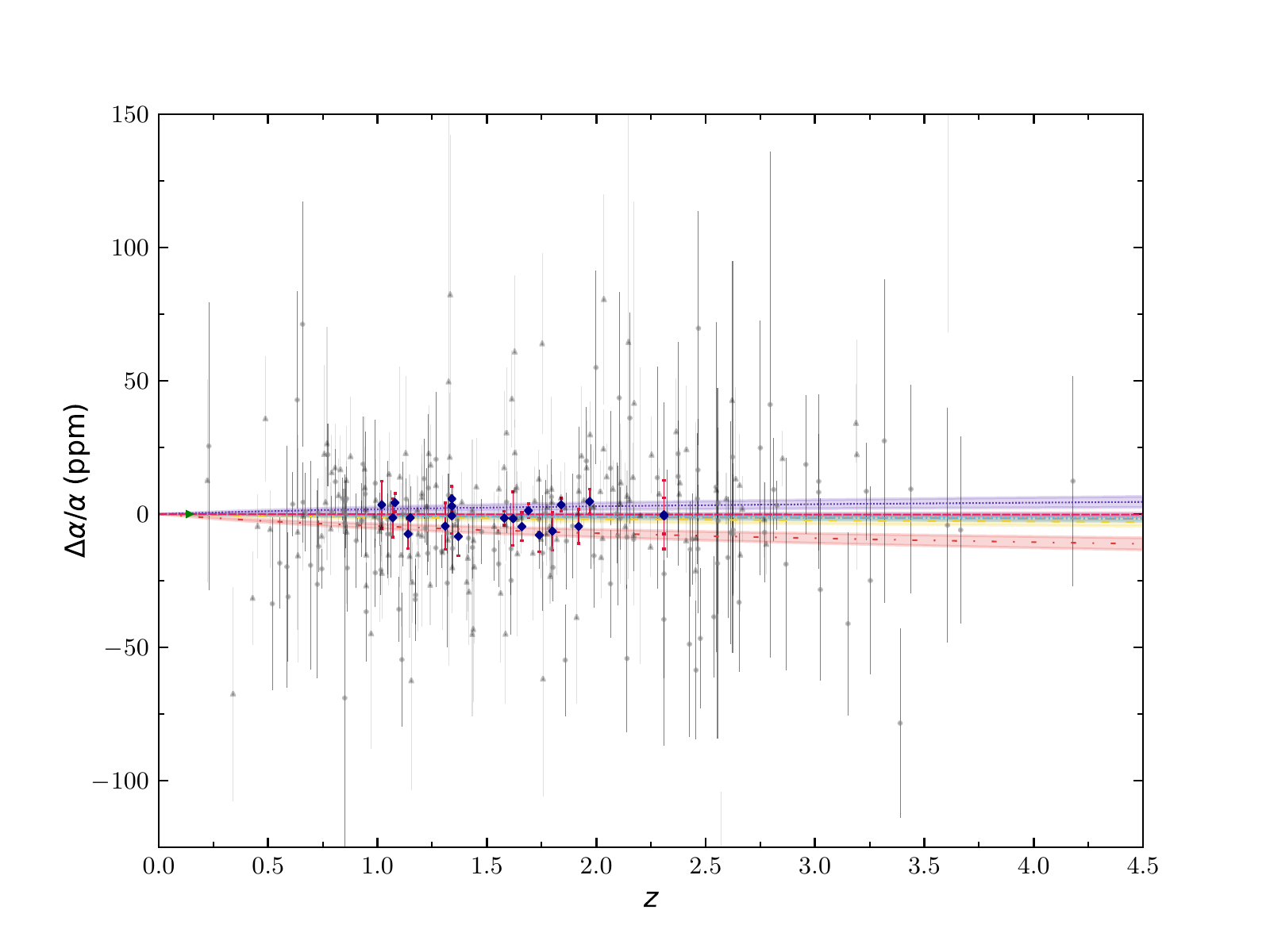}
\end{center}
\caption{\label{fig:eta_evo_posteriors}{Best--fit $\Delta\alpha/\alpha$ evolution along with the $1\sigma$ posterior spread for the $\eta_0^{}$ (top left), $\eta_1^{}$ (top right), $\eta_2^{}$ (middle left), $\eta_3^{}$ (middle right), and $\epsilon$ (bottom) parametrisations of Eqs. (\ref{eq:eta_0}) -- 
(\ref{eq:epsilon}). The illustrated data points from Keck, VLT, New and Oklo data set measurements are described in section \ref{sec:alpha_data}.}}
\end{figure*}

In our constraint analyses we made use of the currently  available measurements 
of $\Delta\alpha/\alpha$ and $\dot{\alpha}/\alpha\vert_0$. We remark that the constraints on 
the parameters $(\eta_{0,1,2,3}^{},\,\epsilon)$ defining the $\eta(z)$ 
phenomenological functions (\ref{eq:eta_0}) -- (\ref{eq:epsilon}), were 
transposed from the constraints on the variation in $\alpha(z)$ by using the 
direct relation given in Eq. (\ref{eq:alpha_relation}).
We refer to section \ref{sec:f_T_alpha} for a description on the several $\Delta\alpha/\alpha$ constraints, while the adopted atomic clocks laboratory (lab) constraint is specified by
$\dot{\alpha}/\alpha\vert_0=(-1.6\pm2.3)\times10^{-17}\,\mathrm{year}^{-1}$ 
\cite{Rosenband1808}. When we include the latter constraint on the 
temporal variation of the fine--structure constant, we make use of a Hubble constant prior 
likelihood $H_0^R$ 
\cite{Riess:2019cxk}, since we then marginalise over $H_0^{}$ to infer the 
constraints on the $\eta(z)$ parameters.

\subsection{\label{sec:phenom_cur_constraints}Current constraints}

The inferred constraints on the model parameters of the theoretical functions 
defined in Eqs. (\ref{eq:eta_0}) -- (\ref{eq:epsilon}) are reported in Table 
\ref{tab:eta_models}, in which a number of data sets have been adopted as 
indicated in the first column of this table. Although the derived constraints on 
$\eta_{0,1,2,3}$ and $\epsilon$ are all of the order of $10^{-7}$, the NO data 
set along with the laboratory measurement of the current temporal variation in 
$\alpha$, significantly tighten the constraints that are obtained from the Keck 
and VLT data sets. Moreover,  the Keck and VLT data sets are characterised by 
the largest deviation from the distance--duality relation (due to a preference 
for non--null theoretical model parameter values) irrespective from the adopted 
parametrisation, as clearly illustrated by their joint posterior distribution in 
Fig. \ref{fig:eta_mar_posteriors}. Additionally, the inferred constraints from 
the Keck and VLT data sets lead to incompatible theoretical evolution of 
$\Delta\alpha/\alpha$. This is shown in Fig. \ref{fig:eta_evo_posteriors}, in 
which we depict the best--fit redshift evolution of $\Delta\alpha/\alpha$, along 
with the 68\% uncertainty region.

From the NO, NO + lab + $H_0^R$ and KVNO + lab + $H_0^R$ joint data sets, we 
obtain a minute deviation from the distance--duality relation $(\eta(z)\simeq1)$, 
and such result is independent from the adopted theoretical parametrisation. It 
is worth mentioning that our derived constraints are orders of magnitude more 
restrictive than the ones obtained from cosmological observations, such as in 
Refs. 
\cite{Lv:2016mmq,Costa:2015lja,Ma:2016bjt,Lin:2018qal,Ruan:2018dls,
Holanda:2012at}. We remark that, as expected, the obtained results are in 
agreement with Ref. \cite{Goncalves:2019xtc}, in which the parametrisation 
independence has been further shown with the expected data from upcoming 
experiments. Moreover, we observe that the laboratory measurement is 
complementary to the NO data set, and the inclusion of the laboratory 
measurement did not alter the inferred constraints from the NO data set. Thence, 
in the analyses of section \ref{sec:f_T_constraints}, we exclude the laboratory 
measurement from our data sets, however we have verified that the final results 
do not change.

\bibliographystyle{JHEP}
\bibliography{references}

\end{document}